\documentclass[1p,number,sort]{elsarticle_arXiv}

\usepackage[latin1]{inputenc} 
\usepackage{todonotes}
\usepackage{caption}
\usepackage{lipsum}
\usepackage{float} 

\usepackage{siunitx}

\usepackage{gensymb}

\usepackage{booktabs}

\usepackage{amsmath}

\usepackage{mathtools}
\DeclarePairedDelimiter{\evdel}{\langle}{\rangle}

\usepackage[english,breaklinks=true,colorlinks=true]{hyperref} 
\usepackage[hyphenbreaks]{breakurl} 
\usepackage{xcolor}
\definecolor{c1}{rgb}{0,0,1} 
\definecolor{c2}{rgb}{0,0,1} 
\definecolor{c3}{rgb}{0,0,1} 
\hypersetup{
    linkcolor={c1}, 
    citecolor={c2}, 
    urlcolor={c3} 
}
\usepackage{amsmath,amssymb} 

\usepackage[version=3]{mhchem}

\usepackage{longtable}
\usepackage{supertabular}
\usepackage{amssymb}
\usepackage{amsfonts}
\usepackage{parskip}
\renewcommand{\d}{{\,\rm  d}}

\newcommand{\T}[1]{{#1}^{\sf  T}}

\newcommand{\pd}[2]{\displaystyle\frac{\partial #1}{\partial #2}}

\newcommand{\grad}[1]{{\rm grad}\left( #1 \right)}
\renewcommand{\div}[1]{{\rm div }\left( #1 \right)}

\newcommand{\fsym}[1]{{\rm sym }( #1 )}

\newcommand{\unit}[1]{\rm #1}

\newcommand{\fempty}[1]{{}}

\newcommand{\sty}[1]{\mbox{\boldmath $#1$}}

\newcommand{\fa}{\sty{ a}}
\newcommand{\fb}{\sty{ b}}

\newcommand{\fg}{\sty{ g}}

\newcommand{\fn}{\sty{ n}}

\newcommand{\fq}{\sty{ q}}

\newcommand{\fs}{\sty{ s}}

\newcommand{\fu}{\sty{ u}}

\newcommand{\fx}{\sty{ x}}

\newcommand{\fA}{\sty{ A}}
\newcommand{\fB}{\sty{ B}}

\newcommand{\fI}{\sty{ I}}

\newcommand{\fR}{\sty{ R}}
\newcommand{\fS}{\sty{ S}}

{\begin{equation}\begin{array}{c}}%
{\end{array}\end{equation}}

\usepackage{lineno,hyperref}

\journal{arXiv}

\begin{document}

\begin{frontmatter}

\title{Transient temperature calculation method for complex 
fluid-solid heat transfer problems
with scattering boundary conditions}




\author[mymainaddress]{Peter H\"olz\corref{mycorrespondingauthor}}
\cortext[mycorrespondingauthor]{Corresponding author. Tel.: +49 711 911 88680}
\ead{peter.hoelz@porsche.de}

\author[mysecondaryaddress]{Thomas B\"ohlke}
\author[mymainaddress]{Thomas Kr\"amer}


\address[mymainaddress]{Porsche AG, Porsche Motorsport, Porschestr. 911, 71287 Weissach, Germany}
\address[mysecondaryaddress]{Chair for Continuum Mechanics, Institute of Engineering Mechanics, Karlsruhe Institute of Technology (KIT), Kaiserstr. 10, 76131 Karlsruhe, Germany}

\begin{abstract}
A calculation method for engine temperatures is presented. Special focus is placed on the transient and scattering boundary conditions within the combustion chamber, including fired and coasting conditions, as well as the dynamic heat transfer of the water jacket. Model reduction is achieved with dimensional analysis and the application of probability density functions, which allows for a timescale separation. Stationary in-cylinder pressure measurements are used as input values and, according to the transient behavior, modified with an own part-load model.\\ 
A turbocharged SI race engine is equipped with 70 thermocouples at various positions in proximity to the combustion chamber. Differentiating from already published works, the method deals with the transient engine behavior during a race lap, which undergoes a frequency range of 0.1-1 Hz. This includes engine speed build-ups under gear changes, torque variations, or the transition from fired to coasting conditions.\\ 
Different thermal behaviors of various measuring positions can be simulated successfully. Additionally, cylinder individual temperature effects resulting from an unsymmetrical ignition sequence and different volumetric efficiencies with unequal residual gas can be predicted. Up to a few percent, the energy balance of the water jacket is fulfilled and variations of water inlet temperatures can be simulated accurately enough.
\end{abstract}


\end{frontmatter}

\begin{center}\textnormal{{\scriptsize Copyright, including manuscript, tables, illustrations or other material submitted as part of the manuscript, is assigned to the authors.}}\\[2ex]\end{center}

\tableofcontents
\section{Introduction}
\label{intro}

\subsection{State of the art}
 
A precise engine thermal management should be targeted at as little as possible heat transfer to solid components. The underlying thermomechanical fatigue mechanisms are very sensitive to temperature changes. As an example, aluminium alloys like wrought alloy 2618A show a great temperature dependence of its high cycle fatigue resistance \citep{Robinson2003a}. Different fatigue mechanisms like low cycle fatigue (LCF) or high cycle fatigue (HCF) react differently to temperature changes. Furthermore, other fatigue mechanisms like creeping become more important with increasing solid temperatures. The well-known deformation and fracture mechanism maps \citep{AshbyDefo} can give a good overview in each case. In order to describe quantitatively crack initiation and propagation caused by the combination of all possible fatigue mechanisms, calculation methods are required to determine solid temperatures.\\
Transient engine behaviour and the resulting temperatures also influence other aspects like fuel consumption or emissions. A new method for estimating transient engine-out temperatures and emissions is presented in \cite{Gao2009}. As an example, an increase of fuel consumption about 25 percent is reported at a cold start with a gasoline-powered engine. Another example are HC emissions in SI engines: about 60-80 percent of the emissions result in the cold initial phase when the engine, especially the catalyst, is cold \citep{Heck2001}. One part of the problem arises from flame extinction at cold walls. However, due to re-diffusion into the extinguishing flame, some unburnt hydrocarbons can be consumed \citep{Warnatz2006}. One possibility to deal with cold initial phases are hydrocarbon trap strategies \citep{Heck2001}: cold HCs are adsorbed until the TWC catalyst reaches the lightoff temperature.\\
Early works addressed the problem of engine heat transfer with dimensional analysis and experimental studies: \cite{Eichelberg1939}, \cite{Woschni1967}, \cite{Hohenberg1979}. In addition to these more phenomenological results, simple lumped turbulence modeling techniques can show the physical background of heat transfer in more detail: using a global $k$-$\varepsilon$ model, \cite{Schubert2005} proposed a heat transfer model with a Reynolds-Colburn analogy. In chapter \ref{Fullload_condition}, a more detailed description of quasi-dimensional turbulence modeling is given. There also exist many works which uses three-dimensional CFD in-cylinder flow simulations, including combustion and heat transfer processes: \cite{Payri2005}, \cite{Mohammadi2008a}, \cite{Mohammadi2010a}. \cite{Chiodi2001} proposed a method which couples detailed CFD techniques with a simplified engine working process analysis in order to ensure the overall heat transfer rate. Additionally, specific models were developed for different kind of engines or flow structures: Relating to heat transfer, HCCI (Homogeneous Charge Compression Ignition) engines are investigeated in \cite{Broekaert2017HCCI}, \cite{Soyhan2009} and \cite{Chang2004}. Similarly, hydrogen engines are studied in \cite{Michl2016}, \cite{Demuynck2011} or \cite{Shudo2000}. In this context, a design of experiments method is applied in \cite{Broekaert2016II}: Various engine settings like, e.g., ignition timing, air-fuel ratio, fuel or compression ratio, are investigated. In a more fundamental manner, the Polhausen equation in seven different operating regimes is verified. \\
In literature, one can find various stationary temperature analysis of combustion engines by using CFD-CHT methods: \cite{Fontanesi2013Comb}, \cite{FONTANESI_Boiling}, \cite{Fontanesi2011}. Unfortunately, steady-state engine temperatures are significantly different from transient conditions. Within the framework of thermo-mechanical fatigue analysis, many transient FEM simulations can be found: \cite{Su2002}, \cite{Metzger2014}, \cite{Seifert2009}, \cite{Nicouleau2002}. However, such typical time sections of these self-contained temperature cycles are in the range of one minute, e.g., 0.02 Hz: run-stop conditions should be simulated. There also exist combustion-cycle resolved thermal analysis with detailed in-cylinder heat transfer treatement: \cite{Mohammadi2008}, \cite{Esfahanian2006}, \cite{Kenningley2012a}. However, these simulations describe a high-frequency timescale in the range of the combustion period, e.g, 8-80 Hz, depending on the engine speed. Crank angle resolved calculations correspondingly deal with frequencies of up to several tens of Kilohertz. In this context, big challenges are transient boundary conditions under transient drives with frequencies between the aforementioned ones, e.g., in the range of 0.1-1 Hz. Two- and three-dimensional transient finite-element models are presented in \cite{Rako1996}. It becomes clear that the correct determination of thermal boundary conditions is mandatory. The essential role of engine operational transients, e.g.,  sudden changes in speed and/or load, is presented in detail. However, continuously changing engine states like transitions between fired and motored condition, including speed build-ups with gear changes and part load sections, are not presented. That is exactly the focus of the presented paper.

\subsection{Outline of the paper}

Regarding daily development work and the current available computing power, it is not possible to investigate every possible thermodynamic state of an engine during a transient drive with detailed 3D-CFD simulations, containing heat transfer treatments. Therefore, the research question can be formulated as follows: With regard to the dynamic response of engine component temperatures and the heat flux of the water jacket, is it possible to develop a transient calculation method which is a workable solution technique in the industrial practice? In particular, with respect to the different thermal behavior of various components and to variations in the engine setting like air-fuel ratios or water temperatures, how should the method be established in order to resolve the aforementioned frequency range of 0.1-1 Hz? \\
Using dimensional analysis and a statistical description of all relevant quantities, a MATLAB\textsuperscript{\textcopyright} - StarCCM+\textsuperscript{\textcopyright} interface was developed and implemented, which provides transient thermal boundary conditions for a subsequent three-dimensional finite volume simulation. By using in-cylinder pressure measurements under stationary conditions as an input, the method can account for different engine mappings or cylinder individual effects. Small deviations from these stationary states under a transient drive, resulting from diverse variations in ignition time or air-fuel ratio, are modelled with an own developed \textit{part load model}, which is described in chapter \ref{PartLoadChapter}. \\
Due to restricted calculation time, a detailed CFD simulation of the water jacket, including turbulence modeling, can not be performed under transient conditions. To calculate water heat transfer, a new method for determining heat transfer coefficients is, therefore, presented.\\
The method was validated with a turbocharged SI engine. Cylinder individual heat transfer phenomena, resulting from different volumetric efficiencies and unequal residual gas, are shown. Moreover, different transient behaviours of various measuring points around the combustion chamber are presented. Finally, a variation of the water inlet temperature and the resulting effect on solid temperatures is shown.

\subsection{Method used in this paper}
\label{Overall_Methodology}

The presented method distinguishes between \textit{inner} and \textit{outer} boundary conditions for an engine. The outer boundary conditions can be defined as a time-dependent, five-dimensional engine state matrix:

\begin{equation}
\underline{M}(t)= \left( n_{\text{engine}}(t),m_{\text{air}}(t),t_{\text{int}}(t),T_{\text{i}}(t),m_{\text{fuel}}(t) \right).
\label{OverviewFiveDimensions}
\end{equation}

The entries are engine speed, air mass flow rate, inlet air temperature, induced
torque and fuel mass flow rate, respectively. Either, these quantities can be measured, or they can be modelled within the electronic control unit (ECU). In the context of a threedimensional, transient finite-volume simulation, one has to determine the inner boundary conditions; more specifically, the thermal boundary conditions. In the following, a method is suggested how one can translate \textit{outer, concentrated} boundary conditions into \textit{inner, distributed} boundary conditions. 

\begin{figure}[H]
\captionsetup{width=1.0\textwidth}
\begin{center}
	 \includegraphics[width=1.0\textwidth]{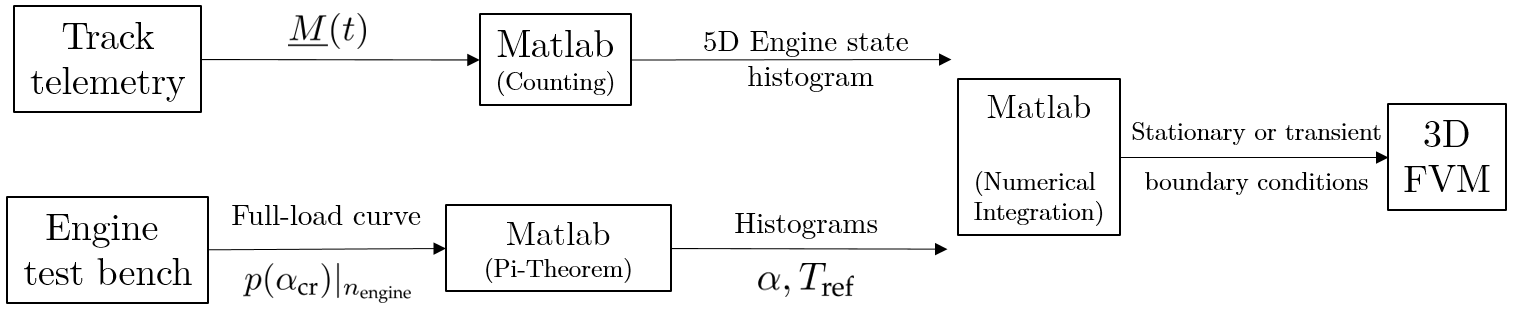} 
\end{center}
\caption{Method overview to generate boundary conditions for a 3D FVM simulation. Track data $\underline{M}(t)$ and pressure indication measurements $p(\alpha_{\text{cr}})$, as function of crank angle $\alpha_{\text{cr}}$, are required. Usually, the reference temperature $T_{\text{ref}}$, which is related to the heat transfer coefficient $\alpha$, is the in-cylinder gas temperture.}\label{fig:Overview_Method}
\end{figure}

An overview of the method is shown in Fig. \ref{fig:Overview_Method}. With the help of an self-programmed code in MATLAB\textsuperscript{\textcopyright}, boundary conditions are determined for a subsequent 3D-FVM simulation of the full engine. Therefore, the engine state $\underline{M}(t)$ is imported by track telemetry with a high enough sampling frequency. All measured states are discretised in a 5D engine state matrix for which a histogram is computed. This histogram is normed in order to use it as an approximation for the probability density function $p_{\underline{n}}$ for the random variable $\underline{n}$ which describes the 5D engine state. Thus, in this case, the engine state is interpreted as a random variable.  For stationary simulations, this density function can be used to get a kind of expectation field of the solid temperature $\evdel{T}(\fx)$. The expectation value with respect to time is denoted by $\langle{\cdot}\rangle$. This field can be used for comparison purposes between different engine mappings or race tracks. It also serves as an initial solution for subsequent transient simulations. Additionally, one needs a high pressure indication measurement $p(\alpha_{\text{cr}})$ of the in-cylinder gas pressure as a function of the crank angle $\alpha_{\text{cr}}$. In this paper, these stationary measurements were resolved with \unit{0.1} {\text{\textdegree}CA}. The full-load curve consisted of 14 engine speed points for which 60 complete engine cycles were recorded in order to take account of the scattering nature of combustion processes. Therefore, a piezo-quartz pressure transducer was mounted on the cylinder head. According to forced convection, thermal boundary conditions are assumed to be of the Newton-form:

\begin{equation}
\fq=- \alpha  \left(  T_{\text{ref}} - T_s   \right)\fn.
	\label{Newton}
\end{equation}

In this case, the heat flux vector $\fq$, which points in the direction of the surface normal vector $\fn$, is calculated from the temperature difference between a well-defined reference temperature $T_{\text{ref}}$ and the solid temperature $T_s$. The proportionality constant is the heat transfer coefficient $\alpha$ (HTC). Different models can be used to calculate this coefficient with the help of the measured pressure data $p(\alpha_{\text{cr}})$. Again, both quantities are interpreted as random variables. However, in this case one has to use conditional probability density functions. In the case of heat transfer coefficients, this is $p_{\alpha|\underline{n}}$ with $\underline{n}$ as the random variable of the engine state. One can easily show that following relation holds for conditional probability density functions \citep{Pope2003}:

\begin{equation}
p_{\alpha | \underline{n}}(A | \underline{N}) = p_{\alpha \underline{n}}(A,\underline{N}) / p_{\underline{n}} (\underline{N}).
\label{PDFConditional}
\end{equation}

The realisations of the both random variables $\alpha$ and $\underline{n}$ are described with $A$ and $\underline{N}$. The joint probability density function is given by $p_{\alpha \underline{n}}(A,\underline{N})$. It must be noted that the two random variables are strongly statistically dependent:

\begin{equation}
p_{\alpha \underline{n}}(A,\underline{N}) \neq p_{\alpha} (A) p_{\underline{n}} (\underline{N}).
\end{equation}

Physically speaking, a heat transfer coefficient $\alpha$ strongly depends on the engine state $\underline{M}(t)$. The description as a random variable is quite useful because one can pay attention to the cyclic fluctuations in the pressure curves as well as the intermittent operation of an engine, without simulating each stroke in detail. Therefore, this method implies a low-pass filter function which only resolves the lower frequencies in the range of the characteristic frequencies of the engine state $\underline{M}(t)$. High frequencies of the individual engine cycles are filtered out. Therefore, in the case of a four-stroke engine, a time discretization in the order of $2 ~ (1/n_\text{engine}) ~ 60 \geq 0.015$ seconds is used.\\
In this paper a race engine is investigated. As can be seen in Fig.\ref{fig:SamplingPoints_LeMans_LoadSpped}, most of the time, such a race engine is in the full-load or in a coasting state. That is the reason why only the full-load curve was measured. However, in chapter \ref{PartLoadChapter}, a part load model is suggested which can be used for modeling probability density functions for engine states which significantly differ from the full-load or coasting state. 
In summary, one gets for any arbitrary random variable $\tilde{f}(\alpha,\underline{n})$ following expressions for its expectation values.\\
In case of a quasistationary simulation in order to calculate a kind of expectation values $\evdel{T}(\fx)$:

\begin{align}
\begin{split}
  \langle \tilde f(\alpha,\underline{n})  \rangle &= \idotsint\limits_{\mathbb{R}^5_{\geq0}} \vspace{55pt}   \int\limits_{\mathbb{R}_{\geq0}}  \tilde f(A,\underline{N}) p_{\alpha \underline{n}}(A,\underline{N})  \,dA  \,d\underline{N}\\
 &= \idotsint\limits_{\mathbb{R}^5_{\geq0}} \vspace{55pt}  \underbrace{\left( \int\limits_{\mathbb{R}_{\geq0}}  \tilde f(A,\underline{N}) p_{\alpha|\underline{n}}(A|\underline{N})  \,dA \right)}_{\langle \tilde f(\alpha,\underline{n}) | \underline{n}=\underline{N} \rangle} p_{\underline{n}}(\underline{N})  \,d\underline{N}.
\end{split}
\label{IntegrationComplete}
\end{align}

In case of a transient simulation in order to calculate time-dependent expectation values $\evdel{T}(\fx,t)$:

\begin{align}
\langle \tilde f(\alpha)  \rangle (t) = \int\limits_{\mathbb{R}_{\geq0}}  \tilde f(A) p_{\alpha|\underline{M}(t)}(A)  \,dA.
\label{TransientIntegrationComplete}
\end{align}

Equation (\ref{IntegrationComplete}) and (\ref{TransientIntegrationComplete}) apply to both, the wetted surfaces with gas, like valves, channels or the combustion chamber, as well as the water channel. However, as will be explained later in chapter \ref{WaterChapterSub1}, the HTC for the water channel is assumed to be clearly determined for a given engine speed $n_{\text{engine}}$. In this case, the probability density function is the Dirac Delta-Distribution. It must be noted that $p_{\alpha | \underline{n}}(A | \underline{N})$ also contains coasting conditions, that means $T_{\text{i}}=0$. The inner integral of equation (\ref{IntegrationComplete}) gives the conditional mean $\langle \tilde f(\alpha,\underline{n}) | \underline{n}=\underline{N} \rangle$ of the random variable $\tilde{f}(\alpha,\underline{n})$. Note that, due to the nonlinearity in equation (\ref{Newton}), a statistically modified reference temperature $\evdel{\alpha T_{\text{ref}} } / \evdel{\alpha} $ is necessary:

\begin{align}
\evdel{\fq} = - \evdel{\alpha}  \left( \frac{\evdel{\alpha T_{\text{ref}}} }{\evdel{\alpha}}  - \evdel{ T_s} \right)\fn.
\label{ExpectationNewtonCorrect}
\end{align}

Whenever it will be spoken of a reference temperature, this modified version is meant. In the following, the acronym ACT, for average cylinder temperature, is used. Remember that equation (\ref{IntegrationComplete}) and (\ref{TransientIntegrationComplete}) also applies to the modified reference temperature. For stationary conditions, details to the proposed statistical method can be found in \cite{PeterSimilar2017}.

\section{Programming details}
\label{ProgrammingDetails}

\begin{figure}[H]
\captionsetup{width=1.0\textwidth}
\begin{center}
	\includegraphics[width=1.0\textwidth]{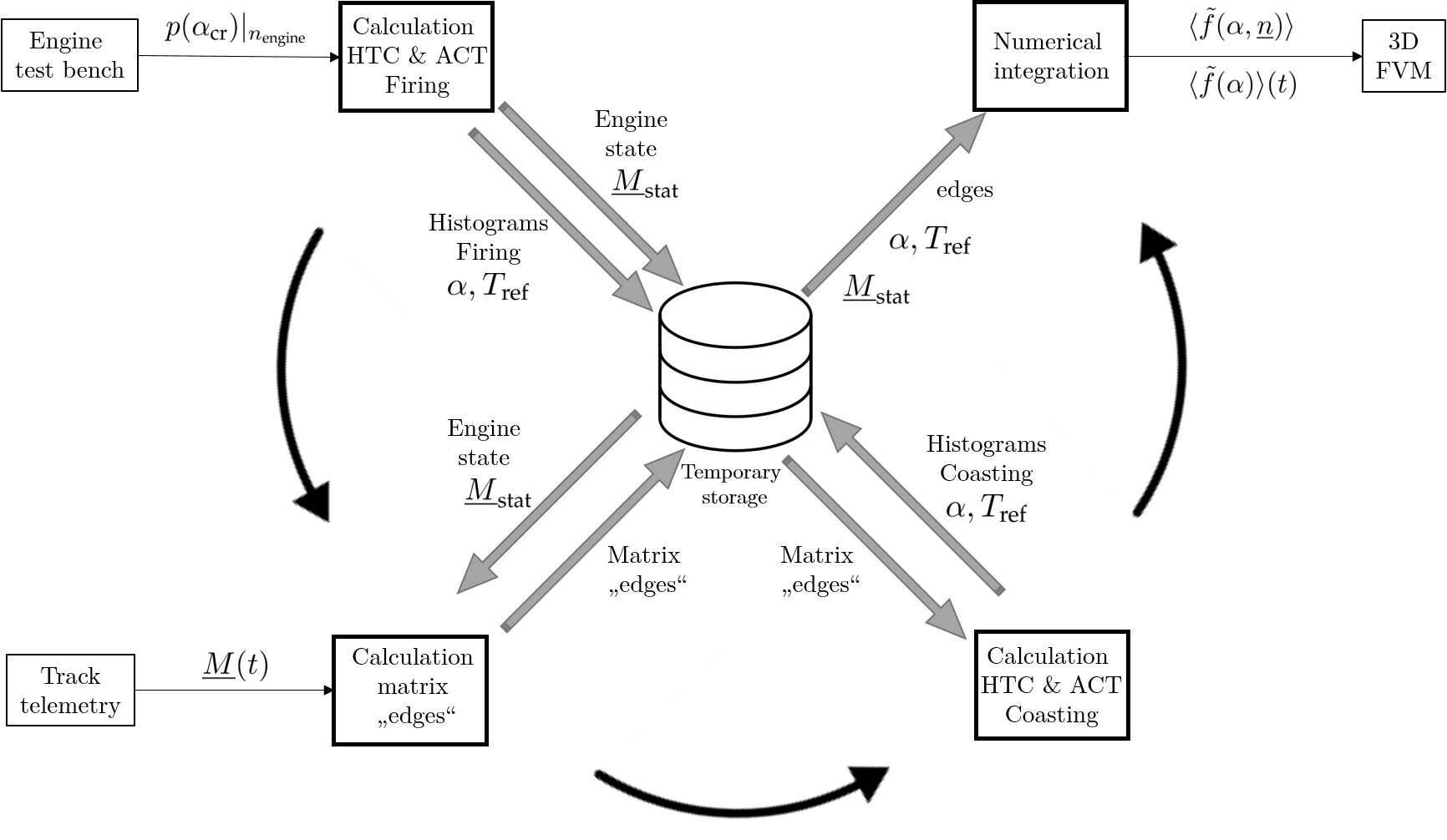} 
\end{center}
\caption{Programming overview.}
\label{fig:Programming_Overview}
\end{figure}

In Fig. \ref{fig:Programming_Overview}, an overview of the code structure is shown. Beginning with the calculation of the probability density functions for the HTC and ACT with the help of pressures measurements, the analysis of the transient engine states follows. The stationary matrix $\underline{M}_{\text{stat}}$ corresponds to the engine state matrix $\underline{M}(t)$ from equation (\ref{OverviewFiveDimensions}) for the stationary measuring points. It serves as a reference state to which the transient states $\underline{M}(t)$ are related. Especially, using the part load model from chapter \ref{PartLoadChapter}, ratios of the matrix entries are of interest.  Afterwards, for each discretized engine state, the coasting condition is modelled and the corresponding probability density functions are calculated. In the end, the PDF's are integrated in order to get quasistationary, $  \langle \tilde f(\alpha,\underline{n})  \rangle$, or transient, $\langle \tilde f(\alpha)  \rangle (t)$, boundary conditions for the subsequent FVM simulation.

The probability density functions are implemented as normed histograms. Therefore, a discretization for each state variable, e.g., the inner and outer boundary conditions, is needed. Just like the modelled HTC and ACT from chapter \ref{Matlabmodeling}, the engine state matrix $\underline{M}(t)$ is represented as a real matrix which is called \textit{edges}. It consists of five dimensions: each of them consists of a vector with different number of elements, depending on the discretization level of the corresponding variable. It is important to note that the histograms have to be normed regarding the corresponding integration field. Multidimensional integration uses Fubini's theorem resulting in a piecewise integration over all dimensions of $\underline{M}(t)$. In case of equation (\ref{IntegrationComplete}), it follows:

\begin{align}
\langle \tilde f(\alpha,\underline{n})  \rangle \approx \sum_{\underline{N}}  \vspace{55pt}  \overbrace{\underbrace{\left( \sum_{A}  \tilde f(A,\underline{N}) \hat{p}_{\alpha|\underline{n}}(A|\underline{N})  \,\Delta A \right)}_{f_{\text{slave}}(\underline{N})} \hat{p}_{\underline{n}}(\underline{N}) }^{f_{\text{master}}(\underline{N})} \,\Delta \underline{N}.
\label{IntegrationCompleteNumerical}
\end{align}

The corresponding normed histograms, with respect to both summations, are described by $\hat{p}_{\alpha|\underline{n}}(A|\underline{N})$ and $\hat{p}_{\underline{n}}(\underline{N})$. Note that the part load model (See chapter \ref{PartLoadChapter} for details) only changes the function $f_{\text{slave}}(\underline{N})$ which has the same structure as the matrix \textit{edges}. It's clear that lots of entries of $\hat{p}_{\underline{n}}(\underline{N})$ are zero because some combinations are not physical: As an example, one can not get a positive value for the induced torque $T_{\text{i}}>0$ if no fuel is injected $m_{\text{fuel}}=0$. One has to mention that the discretization of the engine state matrix $\underline{M}(t)$ can be theoretically infinitely small, depending on calculation power. However, the measured full load line on the test bench has an finite number of engine speed grid points. That is the reason why the function $f_{\text{slave}}(\underline{N})|n_{\text{engine}}$, evaluated for a specific engine speed $n_{\text{engine}}$, is received by linearization between corresponding test bench engine speed grid points $n_{\text{engine}}^{\text{left}}$ and $n_{\text{engine}}^{\text{right}}$: 

\begin{align}
f_{\text{slave}}(\underline{N})|n_{\text{engine}}&=(1-a)f_{\text{slave}}(\underline{N})|n_{\text{engine}}^{\text{left}}+af_{\text{slave}}(\underline{N})|n_{\text{engine}}^{\text{right}},
\end{align}
with
\begin{equation}
n_{\text{engine}}^{\text{left}} \leq n_{\text{engine}} \leq n_{\text{engine}}^{\text{right}},
\end{equation}
and
\begin{equation}
a=\frac{n_{\text{engine}}-n_{\text{engine}}^{\text{left}}}{n_{\text{engine}}^{\text{right}}-n_{\text{engine}}^{\text{left}}}.
\end{equation}

\begin{figure}[H]
\captionsetup{width=1.0\textwidth}
\begin{center}
\includegraphics[width=0.8\textwidth]{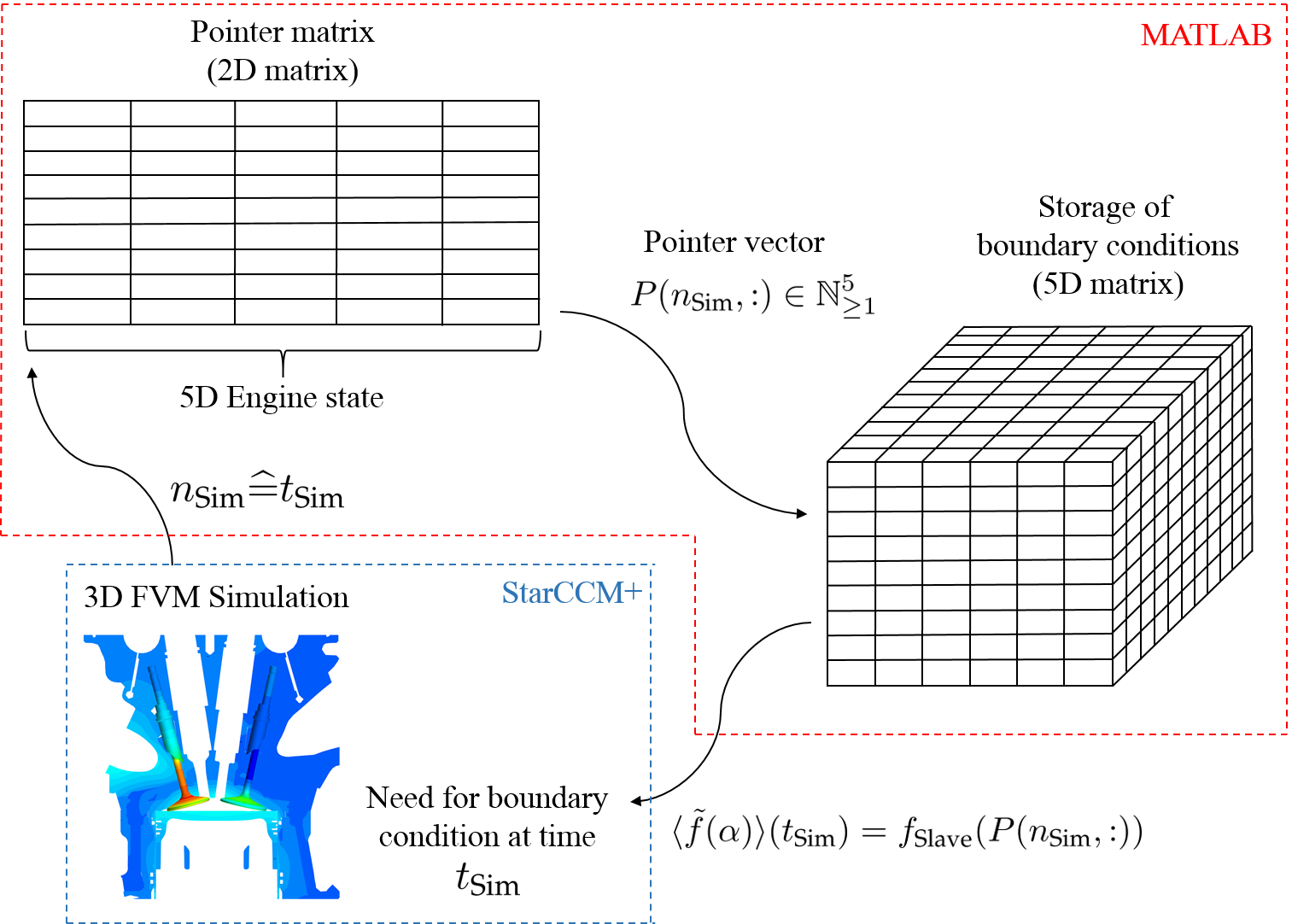} 
\end{center}
\caption{Transient simulation method. Boundary conditions are determined and saved in a MATLAB\textsuperscript{\textcopyright} environment. The subsequent finite volume simulation is done with the aid of the software StarCCM+.}
\label{fig:PointeMethodOverview}
\end{figure}

\begin{figure}[H]
\captionsetup{width=1.0\textwidth}
\begin{center}
	a) \includegraphics[width=0.40\textwidth]{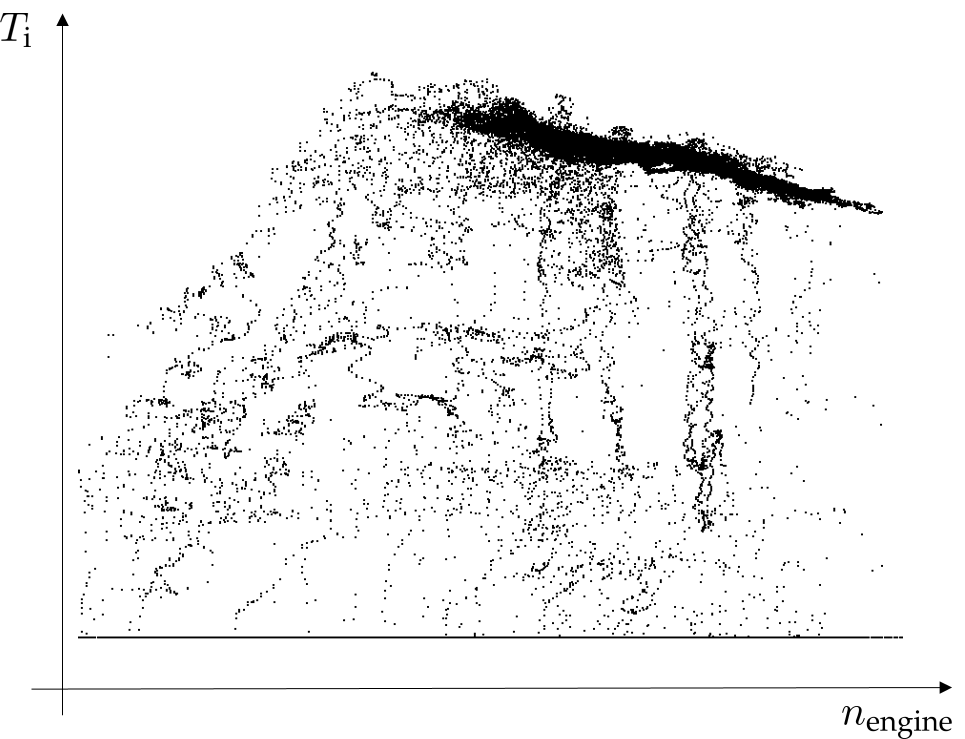} \quad
	b) \includegraphics[width=0.40\textwidth]{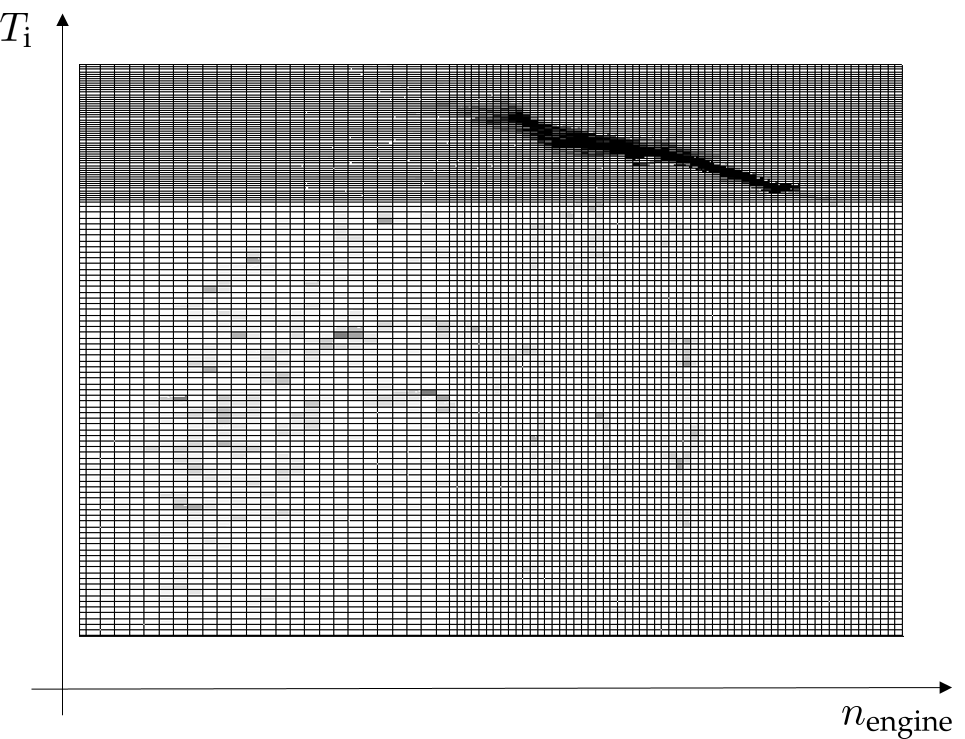}\\
	c) \includegraphics[width=0.45\textwidth]{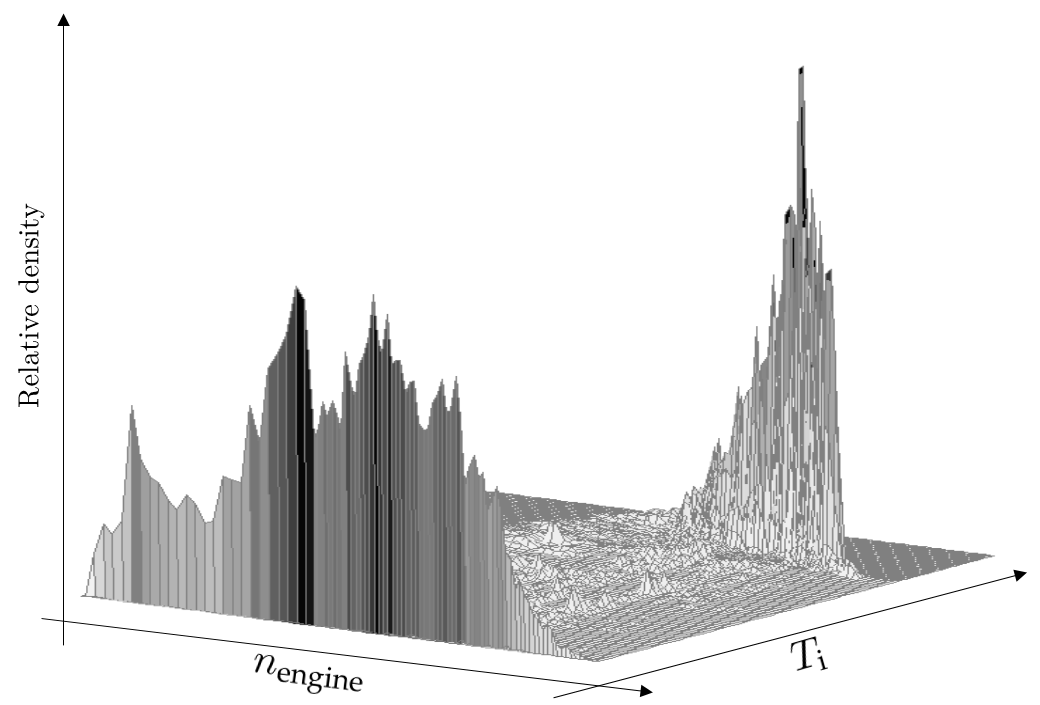}
\end{center}
\caption{a) Exemplary scatter plot in the $T_\text{i}$-$n_{\text{engine}}$ sample space for a typical race lap. b) Corresponding event counting in a discretized sample space with variable mesh size. c) Corresponding relative density of the events.}
\label{fig:SamplingPoints_LeMans_LoadSpped}
\end{figure}

In order to perform transient simulations, one does not need expected values in the form (\ref{IntegrationCompleteNumerical}): in this case, one needs the conditional mean $\langle \tilde f(\alpha,\underline{n}) | \underline{n}=\underline{N} \rangle$. Of course, this value is a function of time because the engine state matrix $\underline{M}(t)$ is a function of time. For this kind of simulation, a pointer matrix $P$ was implemented which is a matrix with the same number of rows as points in time, which depends on the sampling rate of $\underline{M}(t)$ and the temporal discretization of the 3D FVM simulation, and five columns, corresponding to the five-dimensional engine state. These columns contain the position within the matrix $f_{\text{slave}}(\underline{N})$. The procedure is given in Fig. \ref{fig:PointeMethodOverview}: For a given simulation time $t_{\text{Sim}}$, the corresponding matrix row $n_{\text{Sim}}$ contains the pointer vector $P(n_{\text{Sim}},:)$. It provides the necessary indices for the matrix $f_{\text{slave}}$.
As an example, in Fig. \ref{fig:SamplingPoints_LeMans_LoadSpped} a) a projection of the five-dimensional engine state $\underline{M}(t)$ for a typical race lap is shown. The state matrix is projected onto the $T_{\text{i}}$-$n_{\text{engine}}$ subspace. The coasting state is described by the condition $T_{\text{i}}=0$. Statistical analysis shows that about two third of the complete time the engine is in full load state, while about a quarter of the period the coasting condition is fulfilled. In Fig. \ref{fig:SamplingPoints_LeMans_LoadSpped} b) and c), the corresponding discretized sample space is shown. 

\section{Determination of boundary conditions by means of similarity mechanics}
\label{Matlabmodeling}

For all fluid-wetted surfaces, e.g., the combustion chamber walls, the inlet and outlet channels as well as the water channels, boundary conditions like equation (\ref{IntegrationComplete}) or (\ref{TransientIntegrationComplete}) are required. In the following, a brief summary about the most important models is given. 

\subsection{Heat transfer modeling in the inlet and outlet system}

The correlation according to \cite{Caton1981}
\begin{equation}
N\!u_{\text{v}} = 1,84Re_{\text{v}}^{0.58} (D_{\text{v}}/l_{\text{v}})^{0.2}
	\label{OutletInlet}
\end{equation}
is used for the outlet valve stem. The Reynolds number and Nusselt number, which are based on the valve lift $l_{\text{v}}$ and the exhaust jet gas velocity, are described by $Re_{\text{v}}$ and $N\!u_{\text{v}}$, respectively. The diameter of the valve is $D_{\text{v}}$. For the inlet valve stem, a similar expression can be assumed \citep{Yang2000}. In this case, the coefficient in equation (\ref{OutletInlet}) has to be reduced by 40 percent. For the intake port, a simple form of 

\begin{equation}
N\!u = c Re^m,
	\label{SimpleIntakeAndOutlet}
\end{equation}

with a coefficient $c$ and exponent $m$ is used. According to a fully developed flow, in this paper, a value of one was used for the exponent $m$. The coefficient $c$ is used as a calibration parameter. For the exhaust port, following model is proposed \citep{Caton1981}:

\begin{equation}
N\!u=\sqrt{  \left( 8Re_{\text{j}}Pr/\pi \right)}.
	\label{LargeScale}
\end{equation}

In this case, $N\!u$ and $Re_{\text{j}}$ are based on the duct diameter and the exhaust jet gas velocity. The Prandtl number $Pr=\nu/a$ describes the ratio between the kinematic viscosity $\nu$ and the temperature conductivity $a$.

\subsection{Heat transfer modeling in the combustion chamber\\ - Full load condition -}
\label{Fullload_condition}

In this paper, the model according to Bargende is chosen for the heat transfer coefficient $\alpha$ \citep{Bargende1991Original}. 
For the reference temperature in equation (\ref{Newton}), the cylinder-average gas temperature

\begin{equation}
\overline{T}_g=\frac{pV}{NR}
\label{cylinder-average-temperature}
\end{equation}

is used. $N$ is the amount of substance, $R$ the universal gas constant and $V$ the total volume. In the following, only the most important equations for the heat transfer model are given. In complementarity with the original formulation, the characteristic velocity $v$ additionally consists of a scaled combustion convection $v_{\text{c}}$ \citep{BargendeAdditional2011}:

\begin{align}
v&=\sqrt{(8/3)k+v_{\text{p}}^2+v_{\text{c}}^2},\label{CharVeloBargendeI}\\
v_{\text{c}}&= \sqrt[6]{y} ~ \frac{B}{4} \left( \frac{\d y}{\d t} - \frac{T_{\text{ub}}}{\overline{T}_g} \frac{\d x}{\d t} \right) .\label{CharVeloBargendeII}
\end{align}

In equation (\ref{CharVeloBargendeII}), the engine bore diameter is given by $B$. Using the turbulent kinetic energy $k$ and the current piston speed $v_{\text{p}}$, the other two summands describe velocity fluctuations due to turbulence and the in-cylinder flow structure. According to a two-zone combustion model, the ratio between burnt and complete in-cylinder volume is given by $y$. Similarly, the mass fraction is given by $x$. Beginning with the ignition time, the unburnt gas temperature $T_{\text{ub}}$ is calculated with the assumption of a polytropic compression and a homogeneous in-cylinder pressure. According to \cite{Lejsek+Kulzer2010}, the burnt gas temperature is modelled by using the volume balance of the two zones and the ideal gas law. Together with the pressure indication measurements $p(\alpha_{\text{cr}})$ from Fig. \ref{fig:Overview_Method}, the necessary burn function is gained with a pressure trace and combustion analysis.\\
To calculate the turbulent kinetic energy $k$ in equation (\ref{CharVeloBargendeI}), a lumped turbulence model is necessary. In this paper, the model according to \cite{Borgnakke1980_Turbulence} is used. Assuming isotropic, homogenous turbulence for equilibrium conditions, the turbulent dissipation rate $\varepsilon$ is given by $\varepsilon \sim k^{3/2}/l$ with a characteristic eddy length scale $l$, which is given by the combustion chamber volume $V$ according to $l=(6/\pi V)^{1/3}$. On the basis of the rapid distortion theory, e.g., serving the angular momentum, the turbulent kinetic energy $k$ is related to the eddy length scale $l$ according to $k^{1/2} \sim l$. The conservation of mass finally gives the turbulent production rate $\d k \sim 2k/(3\rho) \d \rho$, resulting in the following differential equation:

\begin{equation}
\frac{\d k}{\d t} = - \frac{2}{3} \frac{k}{V} \frac{\d V}{\d t} - \varepsilon_{\text{c}} \frac{k^{3/2}}{l}.
\label{k_DGL}
\end{equation}

Together with the initial condition at the inlet valve closed state, the model constant $\varepsilon_{\text{c}}$ is aligned with a three-dimensional in-cylinder CFD simulation for a representative engine speed of 7000 rpm. The sensitivity of the initial value to the engine speed is then modelled according to \cite{Kozuch2004_Diss}. \\
In \cite{Grill2006_QuasiDim}, within the framework of a quasi-dimensional combustion model, equation (\ref{k_DGL}) is supplemented with a special squish term, and the initial condition is used for engine specified adjustments. Further effects on the temporal evolution of the turbulent kinetic energy, like fuel injection, tumble or swirl flow, are presented in \cite{Grill2006Thesis}. The alternative, quasi-dimensional turbulence model in \cite{LiuJiang2000} contains two differential equations for $k$ and $\varepsilon$. Additionally, the two proportional constants for $\varepsilon \sim k^{3/2}/l$ and $\d k \sim 2k/(3\rho) \d \rho$ are evaluated in more detail. Similarly, \cite{LeeFilipi2011} models a zero-dimensional energy
cascade through a coupled, ordinary differential equation system for the mean kinetic energy and the turbulent kinetic energy. Completing this approach, \cite{DeBellis2014} also considers production terms which are related to the flow through the intake and exhaust valves. For the necessary energy transfer rate, different approaches can be found in literature: \citep{DeBellis2016} or \citep{DeBellis20188II}. The last one systematically transforms the three-dimensional conservation equations of the RANS $k$-$\varepsilon$ model into a quasi-dimensional turbulence model.

\subsection{Heat transfer modeling in the combustion chamber\\ - Coasting condition -}
\label{CoastingChapter}

Using the first three entries of the five-dimensional engine state matrix $\underline{M}(t)$ and an isentropic assumption for the motored pressure during compression and expansion,

\begin{equation}
pV^{\kappa}=p_{\text{ini}}V_{\text{max}}^{\kappa},
\label{IsentropicCompression}
\end{equation}

the heat transfer coefficient under coasting conditions can be easily modelled. In this case, $\kappa$ is the isentropic exponent. The subscripts denotes the initial pressure and the maximum volume.
In literature, on can find various approaches for convective heat transfer equations under motored conditions. The influence of engine speed is investigated in \cite{Sanli2008a}. In \cite{Adrian2013} an alternative model for motored conditions is suggested and compared with conventional engine models. A review about this topic is given in \cite{Broekaert2016I}. The effect of gas properties is explicitly investigated in \cite{Demu2012_Properties}. 

\subsection{Heat transfer modeling in the combustion chamber\\ - Part load condition -}
\label{PartLoadChapter}

\begin{figure}[H]
\begin{center}
	 \includegraphics[width=1.0\textwidth]{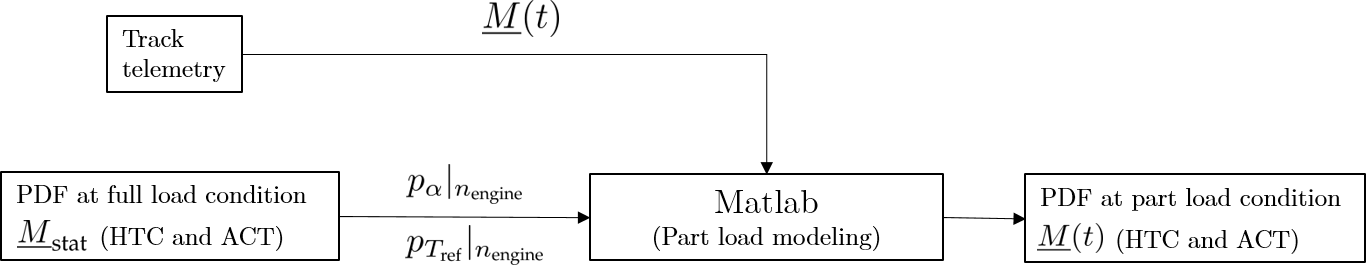} 
\end{center}
\caption{Method overview for generating probability density functions for HTC and ACT at part load conditions. Given are the probability density functions at full load condition and the engine state matrix $\underline{M}(t)$.}
\label{fig:Part_load_modeling}
\end{figure}

Most of the time, a race engine operates in full load or coasting conditions. Nevertheless, a full load state under race conditions differs from stationary measurements on a test bench. According to Fig. \ref{fig:SamplingPoints_LeMans_LoadSpped} a), it does not exist a line, but rather a small subspace of full load points. The ECU has to adjust permanently the operating conditions: Ignition angle, fuel and air mass flow - to name just a few aspects. Moreover, inertia effects of several subsystems like turbochargers cause transient boundary conditions: generating the required boost pressure after coasting needs time resulting afterwards in some overshoots. In the following, an attempt is shown how to model heat transfer coefficients and corresponding reference temperatures with the help of stationary full load data. An overview is given in Fig. \ref{fig:Part_load_modeling}. The probability density functions derived with measured pressure data from the stationary test bench serve as input. In dependence of the transient engine state $\underline{M}(t)$, new probability density functions are modelled. In this paper, all variances from the stationary conditions are called \textit{part load} conditions. 

In the following, a very simple model for the heat transfer coefficient in the combustion chamber is presented. All other heat transfer coefficients can be modelled in an analogous manner. This summary briefly outlines the main techniques: details to the suggested part load model as well as its validation under stationary boundary conditions can be found in \cite{PeterReduction2017}. The matrix entries of $\underline{M}(t)$, describing the transient engine condition according to equation (\ref{OverviewFiveDimensions}), and $\underline{M}_{\text{stat}}$, the corresponding full load state at stationary conditions, are related. Let $\alpha_{\text{stat}}$ and $\alpha(t)$ be two realisations for the heat transfer coefficient in the state $\underline{M}_{\text{stat}}$ and $\underline{M}(t)$, respectively. Using following approach for the transformation of realisations

\begin{equation}
\alpha(t)=\beta(\underline{M}_{\text{stat}},\underline{M}(t),\alpha_{\text{stat}})\alpha_{\text{stat}},
\end{equation}

the new probability density functions according to equation (\ref{IntegrationComplete}) and (\ref{TransientIntegrationComplete}) can be easily calculated. For the mathematical background, see \citep{Pope2003}. 
Starting from the simplified correlation according to Woschni \citep{Woschni1967}

\begin{equation}
\alpha \propto p^m v^m \overline{T}_g^{0.75-1.62m},
\label{Woschni_ratio_simple}
\end{equation}

the transformation coefficient $\beta$ can be described by

\begin{equation}
\beta(\underline{M}_{\text{stat}},\underline{M}(t),\alpha_{\text{stat}})=\left( \frac{p|_{\underline{M}(t)}}{p|_{\underline{M}_{\text{stat}}}} \right)^m   \left( \frac{v|_{\underline{M}(t)}}{v|_{\underline{M}_{\text{stat}}}} \right)^m   \left( \frac{\overline{T}_g|_{\underline{M}(t)}}{\overline{T}_g|_{\underline{M}_{\text{stat}}}} \right)^{{0.75-1.62m}}.
\label{Beta_Ratio_detail}
\end{equation}

\begin{figure}[H]
\captionsetup{width=1.0\textwidth}
\begin{center}
	\includegraphics[width=1.0\textwidth]{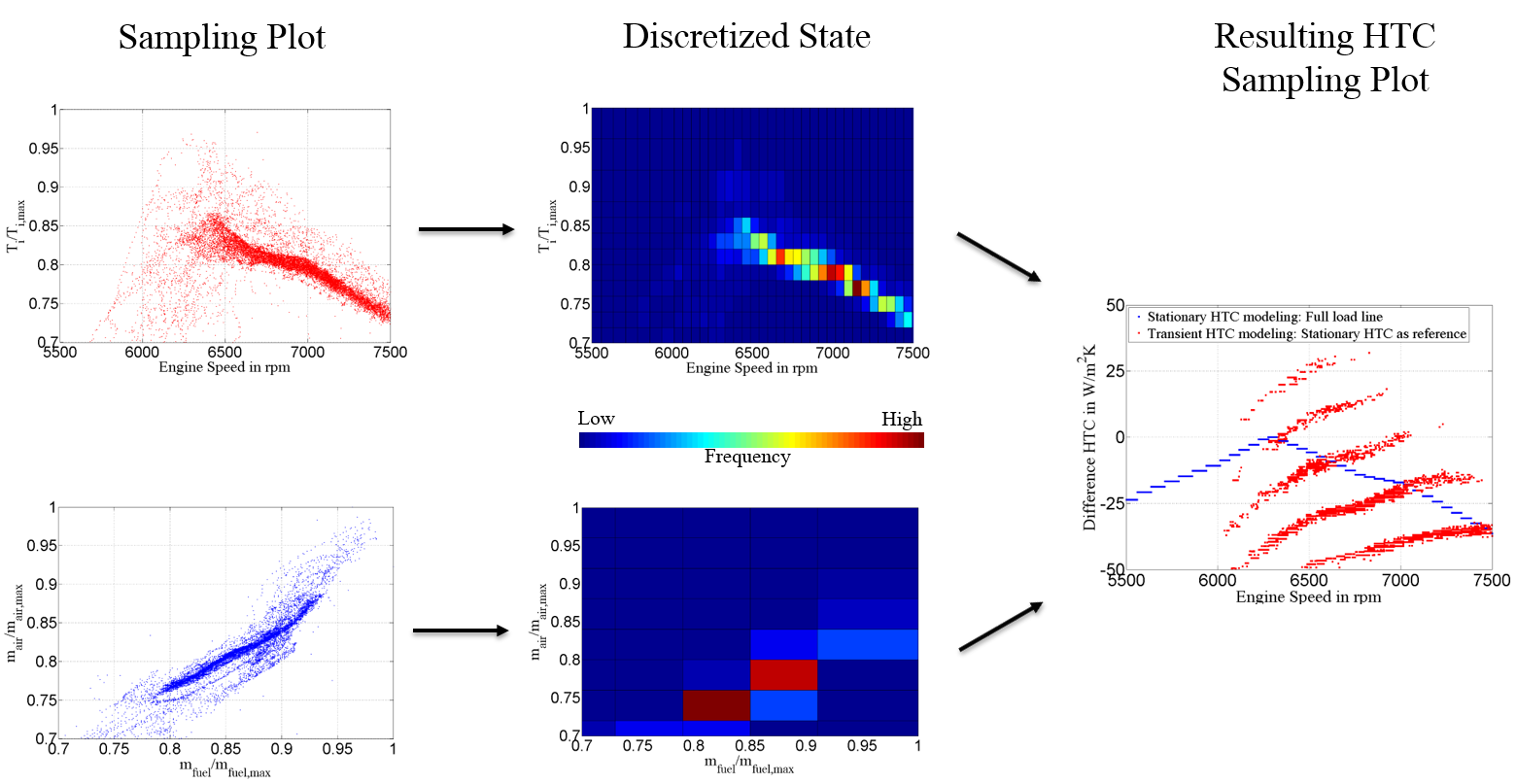} 
\end{center}
\caption{Example for using the part load model for a representative race lap. The $T_\text{i}$-$n_{\text{engine}}$ and $m_\text{air}$-$m_{\text{fuel}}$ sample spaces are shown and discretized with different mesh sizes. Values are normed to corresponding maximum values. The resulting HTC sample space is shown on the right side: Values give the difference to the maximum, stationary HTC.}
\label{fig:PartLoad_Example}
\end{figure}

Approaches for the specific ratios in equation (\ref{Beta_Ratio_detail}), as a function of the matrix entries of $\underline{M}(t)$ and $\underline{M}_{\text{stat}}$, can be found in \cite{PeterReduction2017}. \\
In Fig. \ref{fig:PartLoad_Example}, an example for using the part load model is shown. The inlet temperature $t_{\text{int}}$ was kept constant, resulting in a four-dimensional engine state matrix $\underline{M}(t)$ from equation (\ref{OverviewFiveDimensions}). In order to visualize the different impacts, the $m_\text{air}$-$m_{\text{fuel}}$ sample space is meshed with a much coarser mesh size than the $T_\text{i}$-$n_{\text{engine}}$ sample space. The red separate subdivisions of the heat transfer coefficients result from the coarse mesh. The variation within one subdivision is the result of the $T_\text{i}$-$n_{\text{engine}}$ sample space, which has a finer mesh. For comparison purposes, the full load line, resulting from stationary measurements on the engine test bench, is plotted as a blue reference line. The step function results from the engine speed discretization.

\subsection{Transient heat transfer due to variations of water mass flow and temperature}
\label{WaterChapter}

\subsubsection{Model reduction based on similarity mechanics}
\label{WaterChapterSub1}

According to similarity mechanics, one can assume a correlation of the form $Nu=f(Re,Pr)$ for the water channel. Because of the large heat capacity of water, the temperature in the water jacket varies quite slowly. Its difference during one lap is in the range of \unit{5} {K}. Therefore, the dependence on the Prandtl number $Pr$ can be neglected. With the same argument, the viscosity in the Reynolds number can be ignored, resulting in a pure dependence of the water velocity. Actually, it is very expansive to simulate nearly every occurring water mass flow rate during one lap within a detailed CFD simulation. Additionally, if one wants to investigate solid temperature curves with transient simulations one ought to map the heat transfer coefficients at every time step on the solid mesh, resulting in an extremely high effort. Alternatively, one could simulate the fluid region with the transient Navier-Stokes equations simultaneously. However, due to turbulent flow with thin boundary layers, much more cells would be necessary in comparison with a pure solid simulation. Therefore, a new method is presented by simulating heat transfer by means of CFD for only one reference water mass flow rate. The result of this stationary reference simulation is a heat flux vector for each solid cell, e.g., a field function $\fq_{\text{ref}}(\fx)$. Assuming a correlation $Nu=f(Re,Pr)$ with a Reynolds exponent $m$, following separation approach for the HTC in transient simulations can be formulated:

\begin{equation}
\alpha(\fx,t) \approx \tilde{c}_{\text{ref}}(\fx) v_{\text{water}}^m(t) \sim c_{\text{ref}}(\fx)  n_{\text{engine}}^m(t).
\end{equation}

The proportionality factors $\tilde{c}_{\text{ref}}$ and $c_{\text{ref}}$ contain implicitly the field function $\fq_{\text{ref}}(\fx)$ from the stationary reference simulation. $v_{\text{water}}$ is a characteristic velocity of the water channel flow. Due to the water pump and the fixed transmission ratio with respect to the engine, this velocity is proportional to the engine speed. Consequently, the transient HTC can be approximated with

\begin{equation}
\alpha(\fx,t) \approx \alpha_{\text{ref}}(\fx) \left(\frac{n_{\text{engine}}(t)}{n_{\text{engine}}|_{\text{ref}}}\right)^m.
\label{SeparationApproach}
\end{equation}

$n_{\text{engine}}|_{\text{ref}}$ is the engine speed in the stationary reference simulation. This reference engine speed should be chosen carefully according to the weighted average 

\begin{equation}
n_{\text{engine}}|_{\text{ref}} = \left(  \langle n^m \rangle \right)^{1/m}.
\label{ReferenceSpeed}
\end{equation}

Due to the isothermal assumption in equation (\ref{SeparationApproach}), temperature dependencies of the form $\propto \lambda / \nu^{m} Pr^{n}$ are neglected. In this case, the Prandtl exponent is described with $n$, and the thermal conductivity is denoted with $\lambda$. The kinematic viscosity is $\nu$. The mapped reference heat transfer coefficient $\alpha_{\text{ref}}$ is determined by

\begin{equation}
\alpha_{\text{ref}}(\fx) = \frac{\fq_{\text{cond}}(\fx)\cdot\fn(\fx)}{(T_{\text{ref}}-T_s(\fx))}.
\label{HTC_ref}
\end{equation}

In this case, $\fq_{\text{cond}}$ is the heat flux vector due to heat conduction in the solid part, and $\fn$ is the boundary normal vector. Moreover, $T_{\text{ref}}$ is the specified reference temperature, e.g., the water inlet temperature, and $T_s$ is the solid temperature of the wall next cell. 

\subsubsection{Increased heat transfer due to turbulence}
\label{WaterChapterSub2}

The determination of the reference HTC $\alpha_{\text{ref}}(\fx)$ is based on a CFD-CHT calculation method. Therefore, the SST $k$-$\omega$ turbulence model by Menter \citep{Menter1994} is used. According to the approach by Boussinesq, the Reynolds stress tensor $\fR$ is modelled with the turbulent viscosity $\nu_{\text{t}}(\fx)=k\tilde{T}$ with the time scale $\tilde{T}=\text{max} \left(  \omega,SF_{1}/a_{1}   \right)^{-1}$. $F_{1}$ is a smoothing function which is one at the wall and goes to zero with increasing wall distance. Balance equations are solved for the two additional field variables $k=1/2 \langle{ \fu \cdot \fu  }\rangle$, with the velocity vector $\fu$, and $\omega=\epsilon/k$. According to $\epsilon=2\nu \langle{  \fs \cdot \cdot \fs }\rangle $, the turbulent dissipation is given by $\epsilon$, with $\fs=\fsym{\grad{\fu^{\prime}}}$. The dynamic viscosity is described by $\mu=\nu \rho$, and the density is given by $\rho$. The transport equations for a fixed control volume are then given by

\begingroup
\allowdisplaybreaks
\begin{flalign}
& \frac{\d}{\d t} \int_V \rho k \,dV + \int_A \rho k \sty{ \langle{u}\rangle } \,\cdot d\fa \nonumber \\
\label{k}
& = \int_A \left(  \mu+\frac{\mu_{\text{t}}}{\sigma_{k}}  \right) \grad{k} \,\cdot d\fa + \int_V G_{k}-\rho \beta^* \left(  \omega k \right)  \,dV, \\
& \frac{\d}{\d t} \int_V \rho \omega \,dV + \int_A \rho \omega \sty{ \langle{u}\rangle } \,\cdot d\fa \nonumber \\
\label{eq:kOmegaSTAR2}
& = \int_A \left(  \mu+\frac{\mu_{\text{t}}}{\sigma_{\omega}}  \right) \grad{\omega} \,\cdot d\fa + \int_V  \left[  G_{\omega} - \beta \rho   \omega^2  +  D_{\omega} \right]\,dV. 
\end{flalign}
\endgroup

$\sigma_{k}$ and $\sigma_{\omega}$ describe turbulent Prandtl numbers. $\beta$, $\beta^*$ and $a_{1}$ are model parameters, and $G_k$ is the production term according to $G_{k} = \mu_{\text{t}} S^2$, with $S=\sqrt{2\fS \cdot \cdot \; \fS}$ and $\fS=\fsym{\grad{\langle{ \fu \rangle}}}$. $G_{\omega} = \rho \gamma S^2$ and $D_{\omega} \sim \rho/\omega  \grad{ k} \cdot \grad{ \omega}$ describe additional production terms, whereas $\gamma$ is an additional model parameter. 
Increased heat transfer due to turbulence is calculated with the following expression

\begin{equation}
- \div{\langle{  \fu^{\prime} T^{\prime}  }\rangle} = \div{ \rho c_{\text{p}} a_{\text{t}} \grad{\langle{ T }\rangle} }.
\label{equ_HeatFluxTurbulenceAdd}
\end{equation}

The turbulent Prandtl number $Pr_{\text{t}}$, which describes the ratio between the turbulent viscosity $\nu_{\text{t}}$ and the turbulent temperature conductivity $a_{\text{t}}$, was set to 0.9. The heat capacity for constant pressure is described by $c_{\text{p}}$. Fluctuation values with respect to time are denoted by $\left(\cdot\right)^{\prime}$. Boundary layer velocity and temperature is modelled with the help of the well known wall laws for the viscous and logarithmic layer \citep{Ferziger2002}.

\section{Results and discussion}
\label{Experiment}

\begin{figure}[H]
\captionsetup{width=1.0\textwidth}
\begin{center}
	 \includegraphics[width=1.0\textwidth]{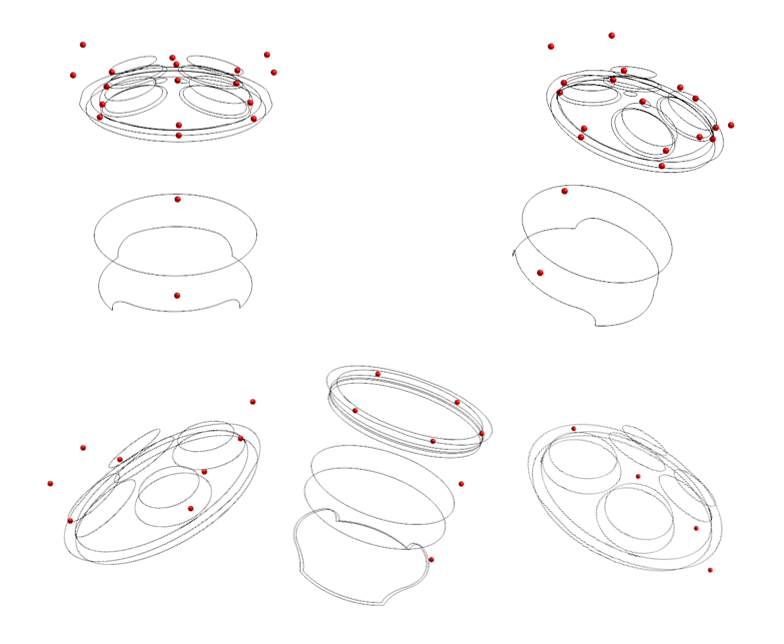} 
\end{center}
\caption{Exemplary measuring points for one cylinder in red. For a clear presentation, some component edges are plotted. Upper row: Overview of all measuring points. Lower row left: Inlet and outlet channels as well as rings. Lower row centred: Liner shoulder with axial variation. Lower row right: Combustion chamber wall of cylinder head with spark plug and intermediate measuring point in between two cylinders.}
\label{fig:MeasuringPointsComplete}
\end{figure}

According to the heat transfer model according to Bargende, the necessary model calibration was done with a single cylinder engine. Afterwards, no modifications of the model parameters were carried out. For model validation, transient measurements on a full engine are used. Both engines are geometrically equal with respect to relevant sections like the water channel, the combustion chamber or the gas channels. 70 measuring points were mounted on a turbocharged engine, which is based on a Porsche racing application. Type K
thermocouples with an uncertainty of \unit{1} {K} were used for the measurement of solid temperatures. A heat-conducting paste with a thermal conductivity of \unit{2.5} {W/(mK)} was put between the solid surface and the thermocouples. Fluid temperatures were measured with Pt100 sensors, which have an uncertainty of \unit{0.053} {K}. Water temperatures were measured at the inlet and outlet. According to equation (\ref{HTC_ref}), these quantities are used as reference temperature for each time step during the transient simulation. Additionally, a measuring turbine for the volume flow rate was installed. In Fig. \ref{fig:MeasuringPointsComplete}, some measuring points for one cylinder are shown. The holes for all measuring points ended about one millimeter below the surface.

\subsection{Single cylinder calibration}

As already mentioned, the HTC models were calibrated with a single cylinder engine. Therefore, one representative, stationary measuring point on the full load line was used: the engine speed was $n_{\text{engine}}=6000$ rpm. 

\begin{table}[H]
\centering
\begin{tabular}{|l|l|l|l|}
\hline
\textbf{Parameter}&\textbf{Value}&\textbf{Parameter}&\textbf{Value} \\
\hline
$\lambda_{\text{cmb}}$ &1.095&max. $m_{\text{fuel}}$ &\unit{80.6} {kg/h}  \\
\hline
$t_{\text{int}}$&\unit{293.15} {K}&$t_{\text{waterin}}$ &\unit{373.65} {K}  \\
\hline
\end{tabular}
\caption{Test parameters for investigation of water heat flow and cylinder individual effects.}
\label{tab:TransientValidation}
\end{table}

\begin{figure}[H]
\captionsetup{width=1.0\textwidth}
\begin{center} 
\includegraphics[width=1.0\textwidth]{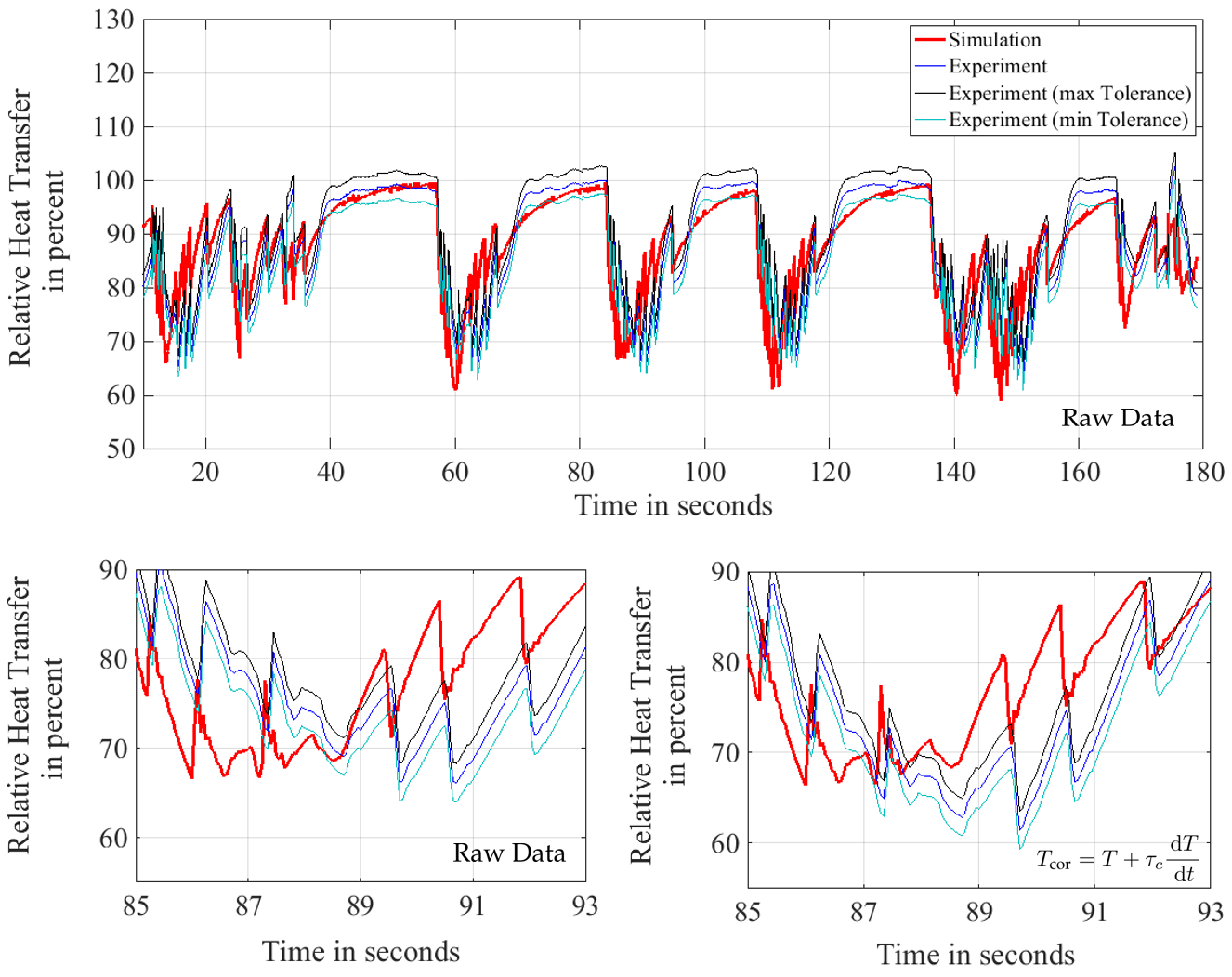}  
\end{center}
\caption{Above: Complete transient water heat transfer for one racing lap. The settling time of about $10$ seconds is not presented. The measurement tolerance takes the uncertainty of PT100 sensors into account. Values are referred to the maximum measured heat flow. Below left: Detail section of the third acceleration phase including some gear changes. Raw data are shown. Below right: Same detail section. Measured temperatures are modified according to $T_{\text{cor}}=T+\tau_c \d{T}/\d{t}$.}
\label{fig:WaterHeatTransferLeMans}
\end{figure}

As an example, a transient drive with the test parameters presented in table \ref{tab:TransientValidation} is investigated. Firstly, the water heat transfer is shown in Fig. \ref{fig:WaterHeatTransferLeMans}. As one can see, the maximum simulated heat flow is within the range of measurement inaccuracy. The heat flow is calculated with the volume flow rate, the heat capacity and the temperature difference between inlet and outlet. In the detail section below, one can see a slightly phase shift within the gear changes. The differences between simulation and experiment is about 10 percent at the beginning. On the left side, raw data are plotted. Because of the thermal inertia effect of sensors, measurement quality can be improved by correcting water temperatures according to $T_{\text{cor}}=T+\tau_c \d{T}/\d{t}$. The time constant $\tau_c$ can be determined by measuring the step response time. However, heat flow measurements have some potential sources of errors: Besides the measurement uncertainty of volume flow rates, the high sensitivity to temperature changes is challenging. Because of the large heat capacity of water, a small temperature change results in a high difference in heat flow. Additionally, the measuring positions of the sensors make some differences. In transient experiments, there is another aspect: strictly speaking, simultaneous measurement of inlet and outlet temperatures is not right. That is the reason for the slightly phase shift in Fig. \ref{fig:WaterHeatTransferLeMans}: the water takes some time to flow through the complete engine. Moreover, pulsating flow can increase or decrease heat transfer in turbulent pipe flows. An analytical study about this problem can be found in \cite{Moschandreou1997} for laminar flows. In addition to the Reynolds number, there are much more system parameters which influence heat flow: Prandtl number, oscillating frequency and amplitude as well as the entry length of pipes. Experimental results with comparable parameters prevailing in engine water channels can be found in \cite{Patel2016} or \cite{Habib2004}. Choosing $Pr \approx 1$, $Re \approx 10^4-10^5$ and the oscillating frequency in the range of $1-5$ Hz, the Nusselt number increase, respectively decrease, is about 10 percent. This can explain some discrepancies during gear changes.

\begin{figure}[H]
\captionsetup{width=1.0\textwidth}
\begin{center}
	 \includegraphics[width=1.0\textwidth]{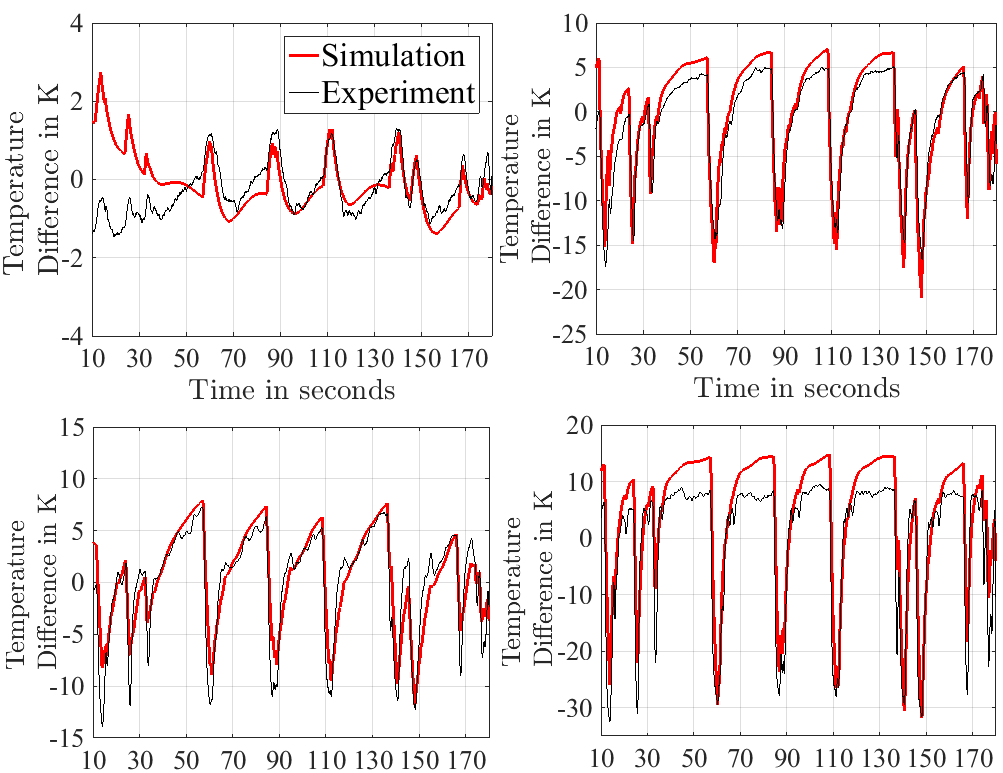} 
\end{center}
\caption{Simulated and measured temperature curves for various measurement positions. As reference temperature, the measured mean values were chosen. Above left: Inlet channel. Above right: Outlet channel. Below left: Inlet ring. Below right: Outlet ring.}
\label{fig:TransientBehaviourAll}
\end{figure}

Secondly, some solid temperature curves are presented in Fig. \ref{fig:TransientBehaviourAll}. Interestingly, one can observe a different thermal behaviour for each engine section. The temperature at the inlet channel is quite constant and the differences between simulation and measurement are negligible small. In the initial phase within the first 30 seconds, the simulation shows a settling process. For the outlet channel, one can observe a strong degressive character for the temperature curve within acceleration phases. The temperature amplitudes are in the order of \unit{20} {K}. The simulation shows weak over- and undershoots of about \unit{3} {K}. However, the mean value is exact enough. The inlet rings show a step-wise linear system response for different time periods. Again, mean values are similar for measurement and simulation. Temperature measurements at the outlet rings are very challenging: Normally, the thermal conductivity is very low resulting in high temperature gradients within a ring. Typical values are about $50$ K/mm. Small position deviations of the thermocouple result in high temperature discrepancies. In this case, it is a strong intrusive measurement method. The high thermal conductivity of thermocouples disturbs the temperature field and increases the heat flow from the outlet rings. Nevertheless, simulated mean values are comparable with measurements. However, the different transient behaviour is noticeable: The simulation shows a continually growing temperature, whereas the measured line is piecewise constant.

\subsection{Cylinder individual effects}

\begin{figure}[H]
\captionsetup{width=1.0\textwidth}
\begin{center}
	 \includegraphics[width=1.0\textwidth]{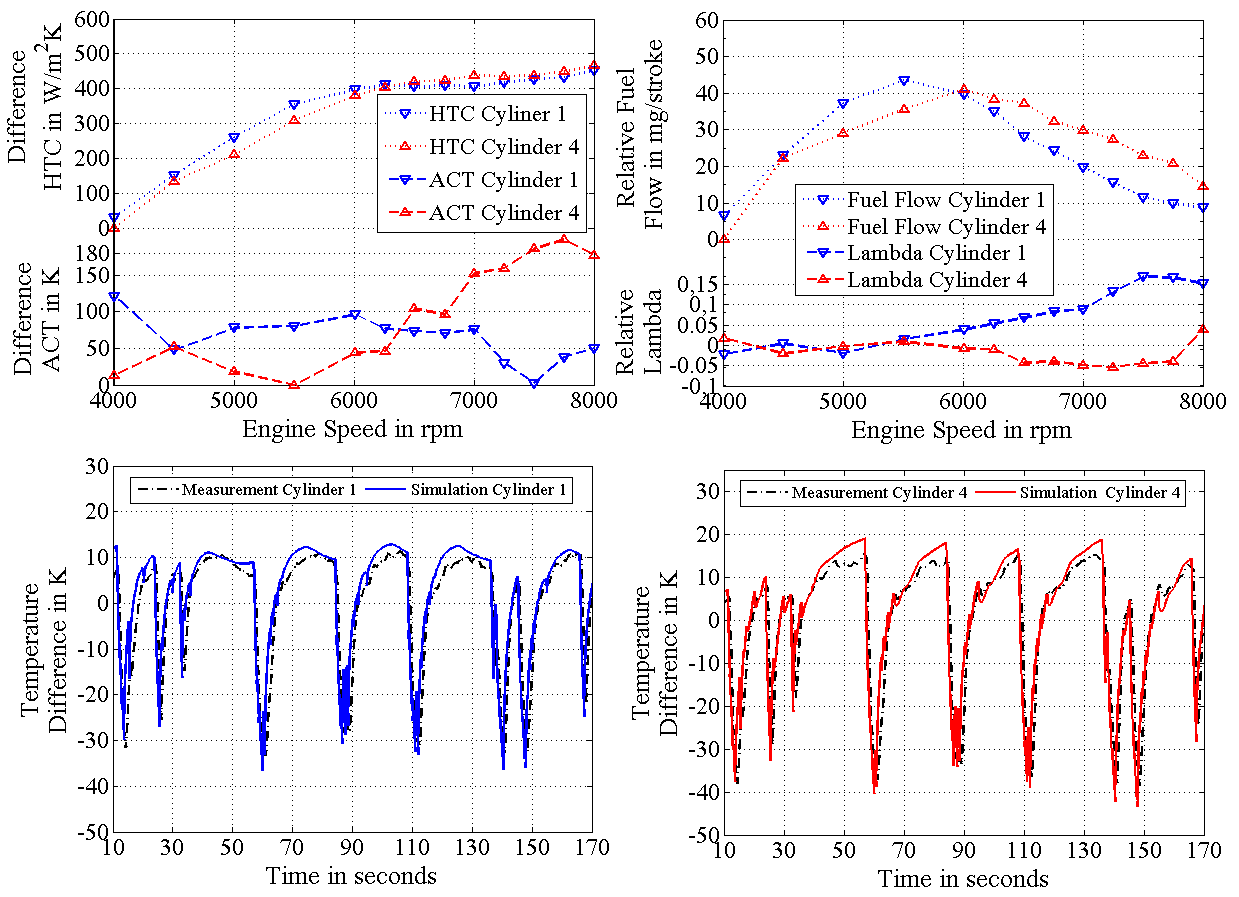} %
\end{center}
\caption{Different thermal loads for cylinder liner shoulders due to individual air-fuel ratios resulting from unequal residual gas. Above right: Injected fuel quantity and lambda value under stationary full load conditions. Lowest fuel flow for cylinder four as well as the mean lambda for both cylinders at \unit{4000} {rpm} are used as reference. Above left: Calculated HTC and ACT for both cylinders under full load state. In each case, the minimum values serve as a reference. Below: Simulated and measured transient thermal behaviour for both cylinders. Temperature differences according to measured mean values are plotted.}
\label{fig:LambdaIndividualBehaviour}
\end{figure}

Because of an unsymmetrical ignition sequence for the investigated engine, there are different volumetric efficiencies with unequal residual gas. As an example, in Fig. \ref{fig:LambdaIndividualBehaviour} two cylinder liner shoulders with different thermal behaviours are presented. 
For both cylinders, mean values as well as amplitudes of the simulated curves are in good agreement with experimental results. Over- or undershoots are temporarily in the range of \unit{5} {K}. It is more interesting that the presented method is able to predict the characteristic curves for each cylinder. Observing an increasing temperature with a subsequent temperature drop at high-speed sections for the first cylinder, one can recognise a continual growth at cylinder four. As can be seen in Fig. \ref{fig:LambdaIndividualBehaviour}, the main reason is the contrary progression of average cylinder gas temperatures with increasing engine speed. Cylinder four has a more stoichiometric combustion, whereas cylinder one gets higher air-fuel ratios resulting in lower gas temperatures. This is physically meaningful because it is well known that the laminar flame speed and the adiabatic flame temperature have their maximum values when $\lambda_{\text{cmb}}$ is a little bit lower than one \citep{Warnatz2006}. Remember that Fig. \ref{fig:LambdaIndividualBehaviour} only shows average gas temperatures. According to equation (\ref{ExpectationNewtonCorrect}), the statistically modified temperature $\evdel{\alpha T_{ref}}/ \evdel{\alpha}$ is needed. For quantitatively accurate simulations, the progression of the HTC is at least as important as gas temperatures. In this case, the qualitative progression is the same for both cylinders. Cylinder one has some slightly smaller values for high engine speed sections.

\begin{figure}[H]
\captionsetup{width=1.0\textwidth}
\begin{center}
	 \includegraphics[width=1.0\textwidth]{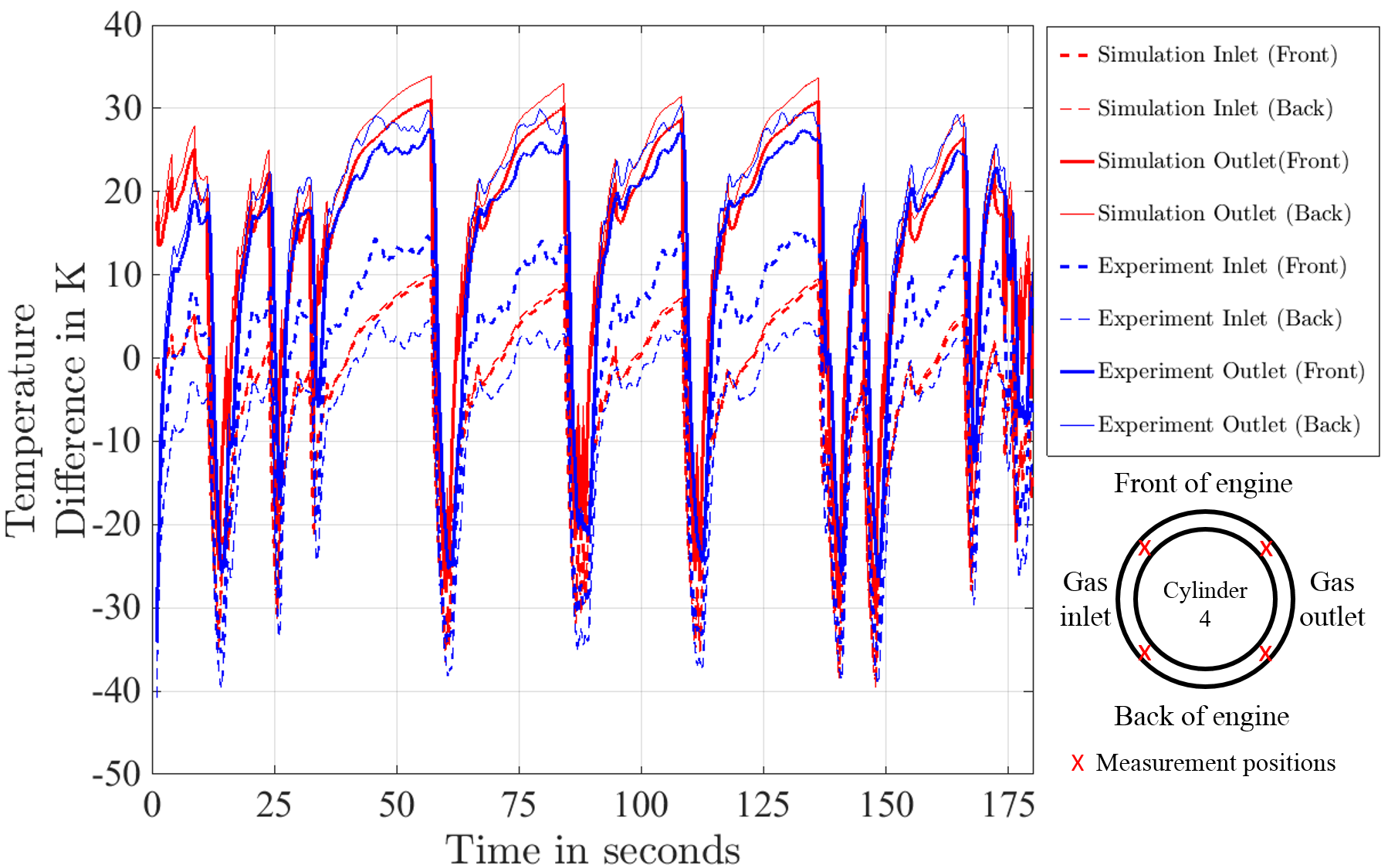} 
\end{center}
\caption{Temperature curves for different measuring positions at the cylinder liner shoulder of cylinder four. The mean temperature of the measured inlet temperature, front of the engine, is used as a reference.}
\label{fig:Liner_Spatial_Cyl4}
\end{figure}

In addition, Fig. \ref{fig:Liner_Spatial_Cyl4} shows some spatial differences at the cylinder liner shoulder. Measuring positions near the exhaust side have about \unit{20}{K} higher temperatures at the end of straight. Main reason is the less heat dissipation because of the hot exhaust channel of the cylinder head. 

\subsection{Variation of water quantities - Simplified calculation method}

\subsubsection{Variation of water mass flow rate}
In order to validate the proposed calculation method for transient water mass flow from chapter \ref{WaterChapter}, stationary measurements with different water mass flow rates are compared with simulation results. Therefore, the engine speed was kept constant. A summary of engine parameters is given in table \ref{tab:WaterVariation}. As a reference water mass flow rate, a value of \unit{2.32} {kg/s} was chosen.

\begin{table}[H]
\centering
\begin{tabular}{|l|l|l|l|}
\hline
\textbf{Engine parameter}&\textbf{Value}&\textbf{Engine parameter}&\textbf{Value} \\
\hline
$n_{\text{engine}}$&\unit{7000} {rpm}&$t_{\text{oilin}}$ &\unit{364.15} {K}  \\
\hline
$t_{\text{int}}$&\unit{310.15} {K}&$t_{\text{waterout}}$ &\unit{361.65} {K}  \\
\hline
$t_{\text{amb}}$ &\unit{301.15} {K}&$m_{\text{fuel}}$ &\unit{100} {mg/stroke}  \\
\hline
$\lambda_{\text{cmb}}$ &1.14&$\alpha_{\text{ign}}$ &\unit{29} {\text{\textdegree}CA} \\
\hline
\end{tabular}
\caption{Engine parameters during variation of water mass flow rate}
\label{tab:WaterVariation}
\end{table}

\begin{figure}[H]
\captionsetup{width=1.0\textwidth}
\begin{center}
	 a) \includegraphics[width=0.45\textwidth]{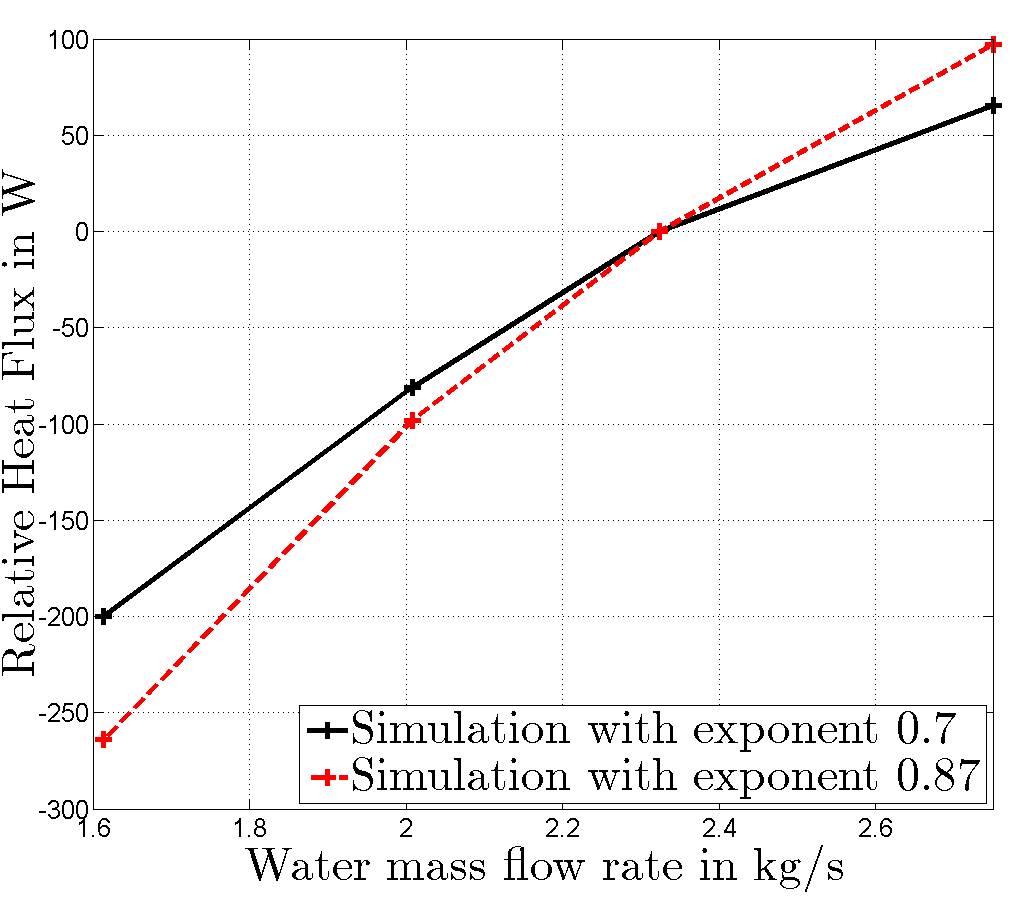} 
	 b) \includegraphics[width=0.45\textwidth]{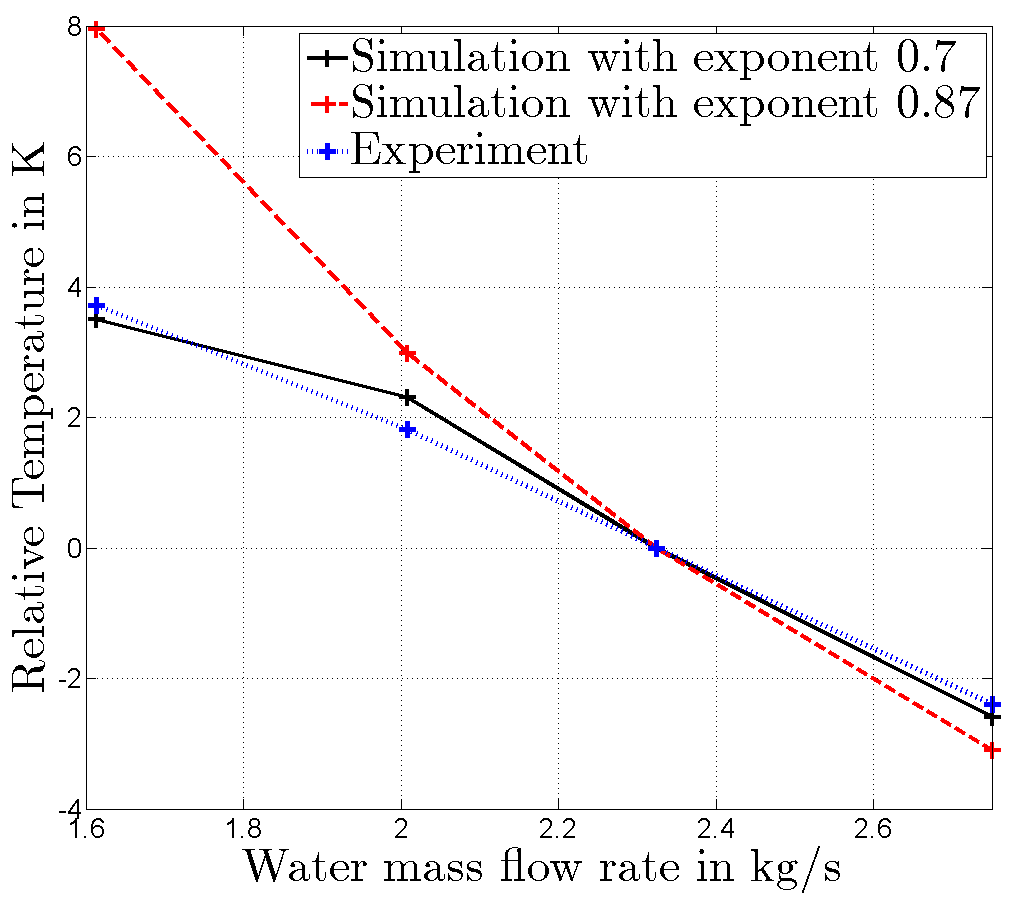}
\end{center}
\caption{a) Relative heat flux to water concerning reference water mass flow rate at one cylinder. Two different Reynolds number exponents for equation (\ref{SeparationApproach}) are shown. b) Resulting relative temperatures concerning reference water mass flow rate at the exhaust valve ring.}
\label{fig:WaterTemperatureAndHeatFlux}
\end{figure}

In Fig. \ref{fig:WaterTemperatureAndHeatFlux}, relative differences in the heat flux and the exhaust ring temperature are shown for two different model parameters $m$ in equation (\ref{SeparationApproach}). In literature, different values can be found: \cite{Taler2017} investigated different forms for the Nusselt correlation in simple turbulent tube flows. They showed that the exponent strongly depends on the Prandtl number. Increasing Prandtl number results in higher exponents. Water temperatures about \unit{373} {K} are typical in engine applications. Using the proposed model in \cite{Taler2017} values slightly larger than 0.8 should be expedient. In comparison with this model, experimental data by \cite{Eiamsa-Ard2010} show only a little bit lower values for the Nusselt number gradient. However, both results are based on a tube length of \unit{1250} {mm}. \cite{Uchida2005} showed that the Reynolds exponent $m$ gets smaller with shorter tube length. This is in accordance with the results presented in this paper. Much better results can be achieved by using Reynolds number exponents in the range of 0.7 instead of 0.8. Because of the complex geometry of engine water channels, models based on shorter tubes seems to be more expedient.   

\begin{figure}[H]
\captionsetup{width=1.0\textwidth}
\begin{center}
	 \includegraphics[width=0.9\textwidth]{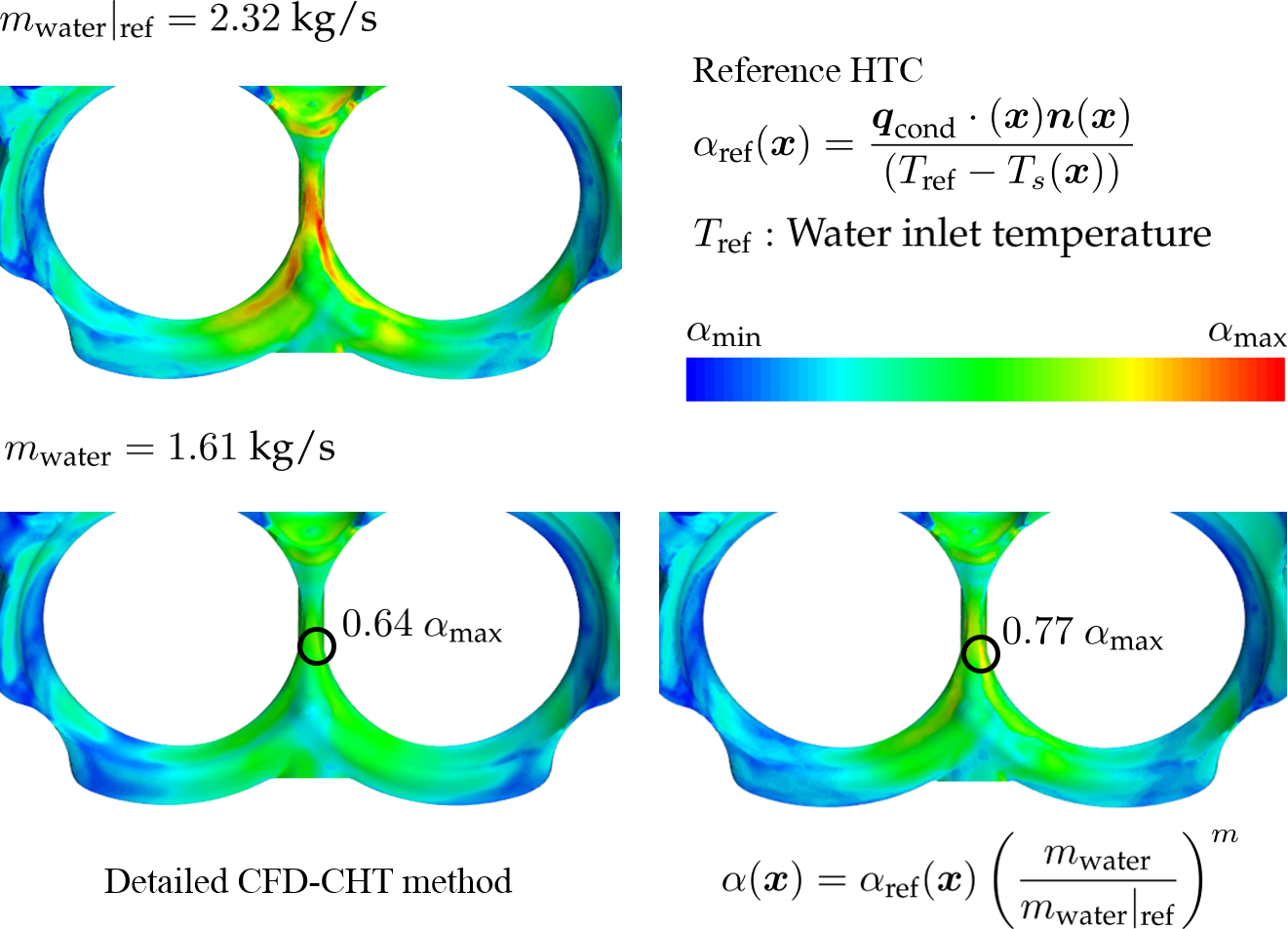} 
\end{center}
\caption{Water channel around the exhaust valve seats: Comparison of the field function $\alpha(\fx)$ for a water mass flow rate of \unit{1.61} {kg/s}. Above: Reference HTC for a water mass flow rate of \unit{2.32} {kg/s}. Lower left: HTC field resulting from a detailed CFD-CHT method with $k-\omega$ turbulence model and wall laws for velocity and temperature. Lower right: HTC field resulting from simplified method according to equation (\ref{SeparationApproach}). The Reynolds exponent $m$ was set to $0.7$. Values give the relation to the maximum value of the reference simulation.}
\label{fig:ComparisonAlphaRefTInlet1}
\end{figure}

Higher values for the Reynolds number exponent result in a higher sensitivity to the relative heat flux concerning the reference water mass flow rate. In Fig. \ref{fig:WaterTemperatureAndHeatFlux} a), the difference of heat flux to the water channel is shown for two simulations with values $m=0.7$ and $m=0.87$. The resulting solid temperatures are shown in Fig. \ref{fig:WaterTemperatureAndHeatFlux} b). As already mentioned, values higher than 0.8 overpredict the sensitivity of the Nusselt number. Using the proposed method from chapter \ref{WaterChapter}, a value in the range of 0.7 seems to be expedient. The corresponding simulated solid temperatures are in good agreement with the experimental results. In Fig. \ref{fig:ComparisonAlphaRefTInlet1}, a comparison of the field function $\alpha(\fx)$, according to equation (\ref{SeparationApproach}), is shown. 

\begin{figure}[H]
\captionsetup{width=1.0\textwidth}
\begin{center}
	 \includegraphics[width=0.9\textwidth]{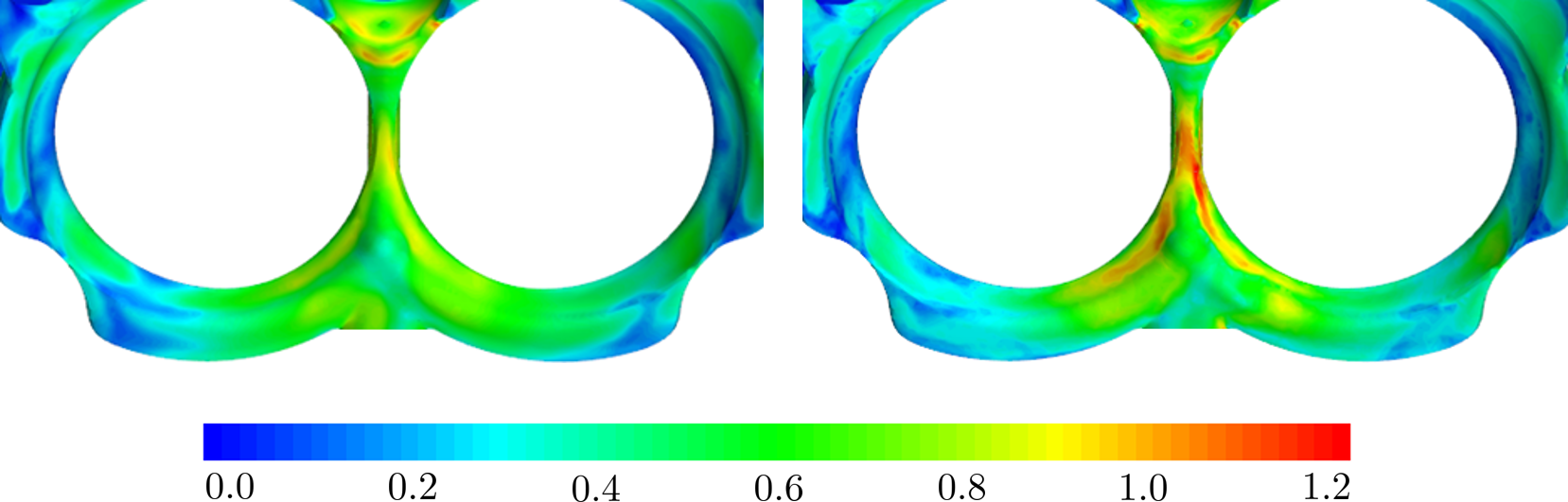} 
\end{center}
\caption{Same plot as Fig. \ref{fig:ComparisonAlphaRefTInlet1}. For a better comparison, values give the relation to the maximum value of the detailed CFD-CHT method for a water mass flow rate of \unit{1.61} {kg/s}.}
\label{fig:ComparisonAlphaRefTInlet}
\end{figure}

Quantitatively, the simplified method overpredicts the HTC values in the intermediate area of the valve seats about 20 percent. However, in some regions, e.g., several small areas on the side walls, the values are a bit lower than the detailed simulation. 

\subsubsection{Variation of water inlet temperature}

\begin{figure}[H]
\captionsetup{width=1.0\textwidth}
\begin{center}
	a) \includegraphics[width=0.45\textwidth]{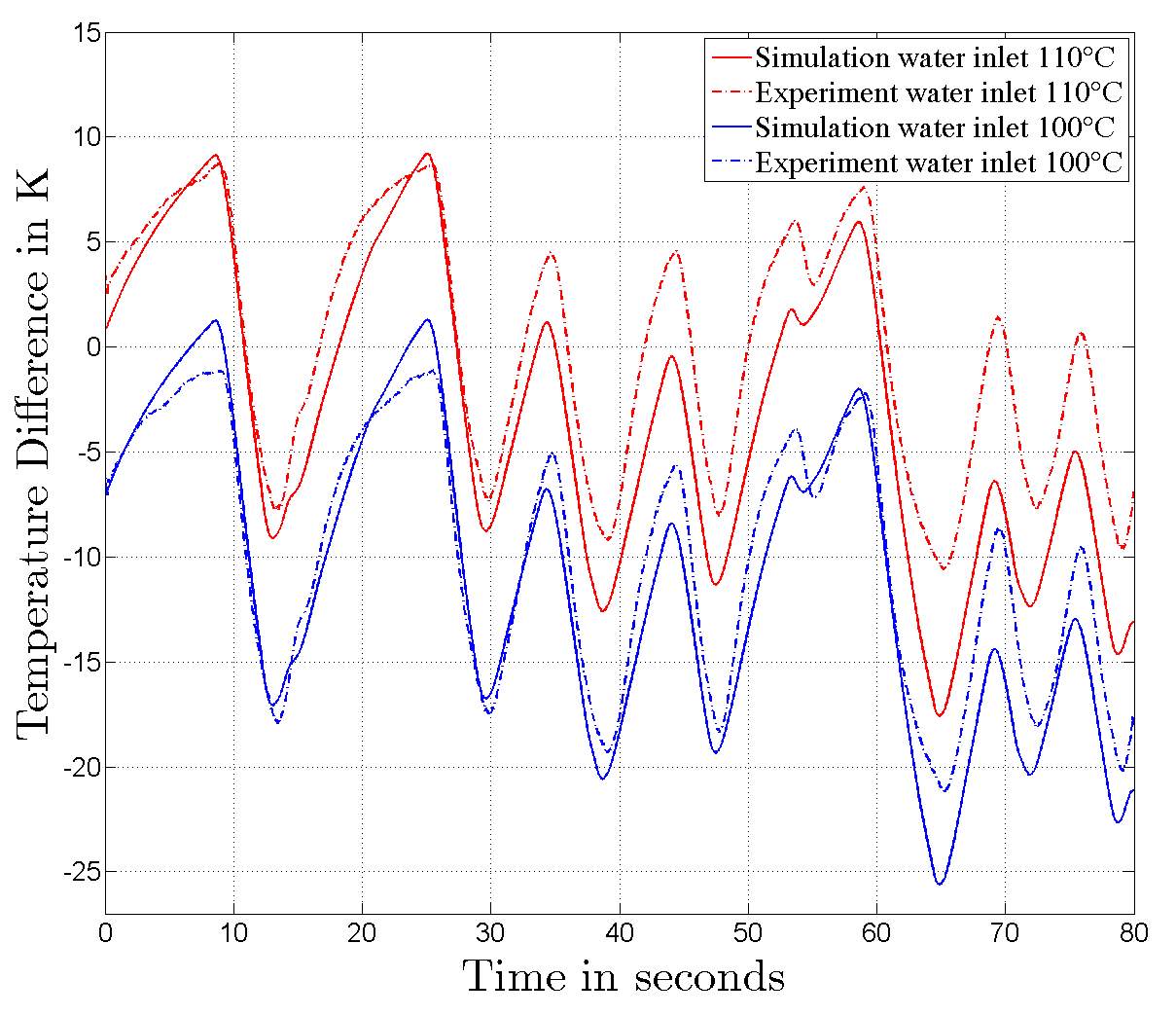} 
	b) \includegraphics[width=0.45\textwidth]{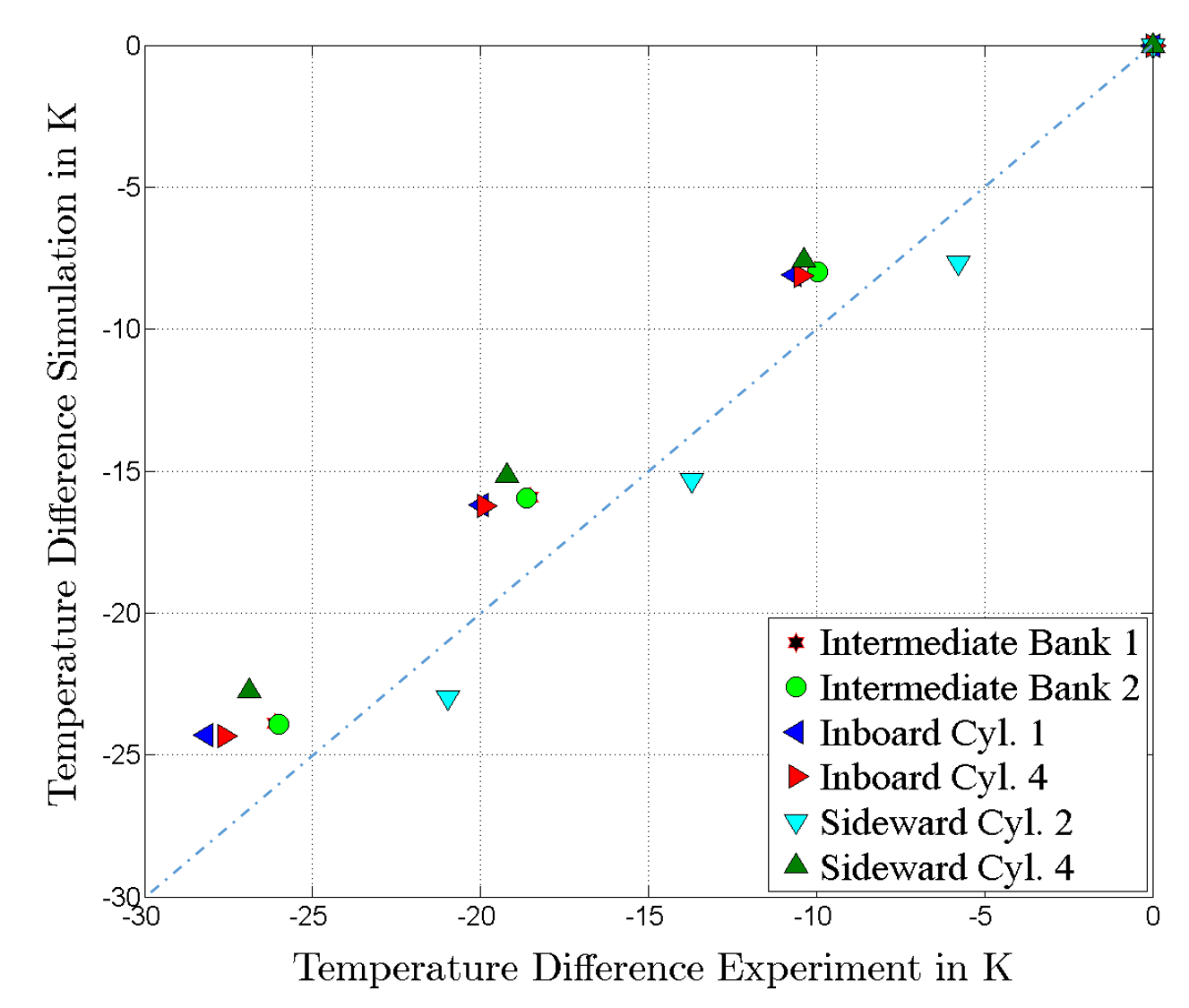}	
\end{center}
\caption{a) Simulated and measured temperature curves for the intermediate measuring point in between cylinder one and two. The mean value of the measured curve ($t_{\text{waterin}}$=\unit{383} {K}) serves as a reference for all four curves. b) Change of mean values for various measuring points at the combustion chamber wall: Inlet water temperature was progressively reduced by \unit{10} {K}. The reference point is $t_{\text{waterin}}$=\unit{383} {K}. The dashed blue lines serve as an orientation and correspond to perfect simulations.}
\label{fig:TWEVariationBarcelona_Intermediate1Point_Relativ}
\end{figure}

Beside the HTC, the reference temperature in equation (\ref{Newton}) should be noted. Therefore, transient measurements with four different water inlet temperatures $t_{\text{waterin}}$ are compared with corresponding transient simulations. Test parameters are given in table \ref{tab:TWEVariation}. As an example, temperature curves for two different water inlet temperatures are shown in Fig. \ref{fig:TWEVariationBarcelona_Intermediate1Point_Relativ} a). As one can see, a change in water inlet temperature results in a kind of temperature offset. In case of Fig. \ref{fig:TWEVariationBarcelona_Intermediate1Point_Relativ} a), this offset is a little bit lower than the change of the water inlet temperature.

\begin{table}[H]
\centering
\begin{tabular}{|l|l|l|l|}
\hline
\textbf{Parameter}&\textbf{Value}&\textbf{Parameter}&\textbf{Value} \\
\hline
$\lambda_{\text{cmb}}$ &1.095&max. $m_{\text{fuel}}$ &\unit{80.6} {kg/h}  \\
\hline
$t_{\text{int}}$&\unit{296.15} {K}&$t_{\text{waterin}}$ & various \\
\hline
\end{tabular}
\caption{Test parameters during variation of water inlet temperature.}
\label{tab:TWEVariation}
\end{table}

Except of the first two acceleration phases, the simulated curves are below the measured lines. In Fig. \ref{fig:TWEVariationBarcelona_Intermediate1Point_Relativ} b), the change of temporal mean values for various measurement points of the cylinder head are shown.

The distinct three collections of dots correspond to a reduction of water inlet temperature from $t_{\text{waterin}}$=\unit{383} {K} to $t_{\text{waterin}}$=\unit{353} {K}. Measurement points at the combustion chamber wall of cylinder one, two and four are shown. A general high sensitivity to water inlet temperature can be observed. However, most of the points are slightly above the dashed blue orientation line. For these points, the simulation underestimates the sensitivity. The intermediate positions of engine bank one and two correspond to points between two cylinders. The inboard points were radially positioned at two thirds of the cylinder diameter. The corresponding sideward points were around the circumference.

\section{Conclusions}
\label{conc}

In this paper, a transient calculation method for complex fluid-solid heat transfer problems is presented. Concerning the initial research question, how should a transient calculation method be established in order to simulate engine temperatures in the industrial practice, the result is as follows: \\ 
Using dimensional analysis and a stationary single cylinder engine calibration, the proposed method can predict solid temperatures accurate enough in the range of 0.1-1 Hz within a reasonable calculation time: Using $80$ CPUs, a complete full engine, which was meshed with $13.5$ million cells for the solid parts, could be simulated within three hours. The corresponding simulation time was one race lap with $180$ seconds. Temperature mean values and amplitudes are in good agreement with experimental data. Different thermal behaviours of various engine components, cylinder-individual temperature effects, as well as the transient heat transfer within the water jacket can be simulated. Consequently, this work complements current simulation techniques in the field of transient thermal analysis of combustion engines in such a way that frequency ranges of 0.1-1 Hz can be successfully simulated.\\
Concerning the water jacket, at the end of straight, the difference between simulation and measurement for the heat flow was within the measurement tolerance. Under transient conditions during gear shifts, the error was less than 10 percent. Differences between the detailed and the simplified method in the local heat transfer coefficient near the valves were found to be in a maximum range of 20 percent. Overestimations, or underestimations respectively, of diverse temperature amplitudes are typically in the range of 2-4 K. Temporally, some values can be up to 7 K.\\
Further investigations, with a wider range of the engine load, should be made. Typical applications are series-production engines with a wide range of part load states.
\newpage
\section{Nomenclature}

\begin{longtable}{l p{0.6\textwidth} r}
	\hline
	\textbf{Symbol} & \textbf{Description} & \textbf{Unit} \\
	\hline
	\endhead
	 \\
	 $a_{\text{t}}$ & Turbulent temperature conductivity  & m\textsuperscript{2}/s\\
	 \\	 
	 $A$ & Realisation of random variable $\alpha$  & W/(m\textsuperscript{2}K)\\
	 \\	 
	 $a_1$ &  Model parameter turbulence modelling & dimensionless\\
	 \\	
	 $B$ & Engine bore & m\\
	 \\ 		 
	 $c_{\text{p}}$ & Specific heat at constant pressure & J/(kgK)\\
	 \\	
	 $edges$ & Matrix which represents $\underline{M}$ & various\\
	 \\	
	 $\tilde{f}$ & Arbitrary random variable & various\\
	 \\
	 $f_{\text{slave}}$ & Approximated inner integral according to equation (\ref{IntegrationCompleteNumerical}) & various\\
	 \\
	 $f_{\text{master}}$ & Argument of outer integral according to equation (\ref{IntegrationCompleteNumerical}) & various\\
	 \\
	 $k$ & Turbulent kinetic energy & (m/s)\textsuperscript{2}\\
	 \\	
	 $l$ & Characteristic eddy length scale & m\\
	 \\ 
	 $m$ & Re exponent & dimensionless\\
	 \\	
	 $m_{\text{air}}$ & Mass flow of air & mg/stroke\\
	 \\
	 $m_{\text{fuel}}$ & Mass flow of fuel & mg/stroke\\
	 \\ 
	 $m_{\text{water}}$ & Mass flow of water & kg/s\\
	 \\ 
	 $\underline{M}$ & Five-dimensional engine state matrix of outer boundary conditions & various\\
	 \\
	 $\underline{M}_{\text{stat}}$ & Stationary engine state matrix of outer boundary conditions for reference purposes & various\\
	 $n$ & Pr exponent & dimensionless\\
	 \\	
	 $n_{\text{Sim}}$ & Matrix row of pointer matrix $P$ for a given simulation time $t_{\text{Sim}}$ & dimensionless\\
	 \\
	 $n_{\text{engine}}$ & Engine speed &  rounds per minute [rpm] \\
	 \\
	 $\underline{n}$ & 5D random variable describing the engine state & various\\
	 \\	
	 $N$ & Amount of substance & mol\\
	 \\		 
	 $\underline{N}$ & Realisation of random variable $\underline{n}$ & various\\
	 \\	
     $\fn$ & Boundary normal vector & dimensionless\\
	 \\	
	 $Nu$ & Nusslet number & dimensionless\\
	 \\	
	 $p$ & Static pressure & Pa\\
	 \\
	 $p_{\alpha \underline{n}}$ & Joint probability density function on $\alpha$ and $n$ & dimensionless\\
	 \\
	 $p_n$ & Probability density function on $n$ & dimensionless\\
	 \\
	 $\hat{p}_n$ & Normed histogram; Approximated probability density function on $n$ & dimensionless\\
	 \\
     $p_{\alpha | \underline{n}}$ & Conditional probability density function on $\alpha$ with regard to $\underline{n}$ & dimensionless\\	 
	 \\
	 $P$ & Pointer matrix & dimensionless\\
	 \\
	 $Pr$ & Prandtl number $\nu/a$ & dimensionless\\
	 \\	
	 $Pr_{\text{t}}$ & Turbulent Prandtl number $\nu_{\text{t}}/a_{\text{t}}$ & dimensionless\\
	 \\	
	 $\fq$ & Heat flux vector & W/m\textsuperscript{2}\\
	 \\	
	 $\fq_{cond}$ & Heat flux vector due to heat conduction in the solid & W/m\textsuperscript{2}\\
	 \\	
	 $R$ & Universal gas constant & J/(molK)\\
	 \\
	 $Re$ & Reynolds number $lv/\nu$ & dimensionless\\
	 \\	
	 $Re_{\text{j}}$ & Exhaust jet Reynolds number $l v_{\text{j}}/\nu$ & dimensionless\\
	 \\		
     $t_{\text{int}}$ & Inlet temperature of air & K\\
	 \\
	 $t_{\text{amb}}$ & Ambient temperature of air & K\\
	 \\	 
	 $t_{\text{waterin}}$ & Inlet water temperature & K\\
	 \\	
	 $t_{\text{waterout}}$ & Outlet water temperature & K\\
	 \\	
	 $t_{\text{oilin}}$ & Inlet oil temperature & K\\
	 \\	
	 $t$ & Physical time & s\\
	 \\
	 $t_{\text{Sim}}$ & Simulation time & s\\
	 \\
	 $T$ & Temperature & K\\
	 \\
	 $T_{\text{ref}}$ & Reference temperature of fluid & K\\
	 \\
	 $\overline{T}_g$ & Cylinder-average gas temperature & K\\
	 \\		
	 $T_{\text{ub}}$ & Unburnt gas temperature & K\\
	 \\
	 $T_s$ & Solid temperature & K\\
	 \\	
	 $T_{\text{i}}$ & Indicated torque by combustion  & Nm\\
	 \\
	 $\fu$ & Velocity vector & m/s\\
	 \\  
	 $v$ & Characteristic velocity & m/s\\
	 \\ 
	 $v_{\text{j}}$ & Exhaust jet velocity through valve opening & m/s\\
	 \\ 	 
	 $v_{\text{p}}	$ & Current piston speed & m/s\\
	 \\ 
	 $v_{\text{c}}	$ & Scaled combustion convection & m/s\\
	 \\
	 $V$ & Volume & m\textsuperscript{3}\\
	 \\
	 $x$ & Ratio between burnt mass and complete in-cylinder mass & dimensionless\\
	 \\	 	
	 $\fx$ & Position vector & m\\
	 \\	 
	 $y$ & Ratio between burnt volume and complete in-cylinder volume & dimensionless\\
	 \\	 	 
\end{longtable}

\section*{Greek symbols}
\begin{longtable}{l p{0.6\textwidth} r}
	\hline
	\textbf{Symbol} & \textbf{Description} & \textbf{Unit} \\
	\hline
	\endhead
	\\
	 $\alpha$ & Heat transfer coefficient & W/(m\textsuperscript{2}K)\\
	 \\	 
	 $\alpha_{\text{stat}}$ & Heat transfer coefficient for stationary engine state $\underline{M}_{\text{stat}}$ & W/(m\textsuperscript{2}K)\\
	 \\	 
	 $\alpha_{\text{ref}}$ & Reference heat transfer coefficient & W/(m\textsuperscript{2}K)\\
	 \\	
	 $\alpha_{\text{cr}}$ & Crank angle & {\text{\textdegree}CA}\\
	 \\
	 $\alpha_{\text{ign}}$ & Ignition crank angle & {\text{\textdegree}CA}\\
	 \\
	 $\beta$ & Model parameter turbulence modelling & dimensionless\\
	 \\
	 $\beta^*$ & Model parameter turbulence modelling & dimensionless\\
	 \\	 
	 $\gamma$ & Model parameter turbulence modelling & dimensionless\\
	 \\		 
	 $\varepsilon$ & Turbulent dissipation & m\textsuperscript{2}/s\textsuperscript{3}\\
	 \\
	 $\varepsilon_{\text{c}}$ & Model constant for turbulent dissipation & dimensionless\\
	 \\
	 $\kappa$ & Isentropic exponent & dimensionless\\
	 \\	
	 $\lambda$ & Thermal conductivity & W/(mK)\\
	 \\	
	 $\lambda_{\text{cmb}}$ & Ratio between actual air mass and stoichiometric air mass & dimensionless\\
	 \\   
	 $\mu$ & Dynamic viscosity & kg/(ms)\\
	 \\	 
	 $\mu_{\text{t}}$ & Turbulent dynamic viscosity & kg/(ms)\\
	 \\	 	  
	 $\nu$ & Kinematic viscosity $\mu/\rho$ & m\textsuperscript{2}/s\\
	 \\	
	 $\nu_{\text{t}}$ & Turbulent kinematic viscosity $\mu_{\text{t}}/\rho$ & m\textsuperscript{2}/s\\
	 \\	
	 $\rho$ & Mass density & kg/m\textsuperscript{3}\\
	 \\	
	 $\sigma_{k}$ & Turbulent Prandtl number for $k$ & dimensionless\\
	 \\	 
	 $\sigma_{\omega}$ & Turbulent Prandtl number for $\omega$ & dimensionless\\
	 \\
	 $\omega$ & Turbulent frequency $\omega=\varepsilon/k$ & 1/s\\
	 \\	
\end{longtable}

\section*{Mathematical Notation}
\begin{longtable}{l p{0.6\textwidth} r}
	\hline
	\textbf{Symbol} & \textbf{Description}& \\
	\hline
	\endhead
	\\	
	 $\grad{\cdot}$ & Gradient: $\pd{\left( \cdot \right)}{u^k} \fg^k$ with contravariant component $u^k$ and corresponding reciprocal basis $\fg^k$. &\\
	 $\div{\cdot}$ & Divergence: $\grad{\cdot} [\fI]$ & \\
	 \\	 
	 $\fI$ & Second order identity tensor, which maps each vector $\fa$ onto itself: $\fI\fa=\fa, ~ \forall \fa$. & \\
	 \\	 
	 $\fa \cdot \fb$ & Scalar product according to $a_{i}b^{i}$ with covariant component $a_{i}$ and contravariant component $b^{i}$. & \\
	 \\	 	 
	 $\fA \cdot \cdot \fB$ & Scalar product according to $A_{ij}B^{ij}$ with covariant component $A_{ij}$ and contravariant component $B^{ij}$. & \\
	 \\	 
	 $\fsym{\fA}$ & Symmetric part of a tensor $\fA$ according to $\fsym{\fA}=1/2\left( \fA+\T\fA \right)$.   \\
	 $\T\fA$ & Transposed tensor of tensor $\fA$ defined by:
	 $\fa \cdot \left( \T\fA \fb \right) = \fb \cdot \left( \fA \fa \right)$ with arbitrary vectors $\fa$ and $\fb$. \\
	 $\langle{\cdot}\rangle$ & Expectation value with respect to time &\\
	 \\	
	 $\left(\cdot\right)^{\prime}$ & Fluctuation value with respect to time & \\
	 \\	
	 \\	 
\end{longtable}

\section*{Abbreviations}

\begin{longtable}[c]{p{0.35\textwidth} p{0.0\textwidth}  r}
	\hline
	\textbf{Abbreviation} & \textbf{} & \textbf{Description} \\
	\hline
	\endhead
	\\
	$ ACT $ & & \textbf{A}verage \textbf{C}ylinder \textbf{T}emperature \\
	$ ECU $ & & \textbf{E}lectronic \textbf{C}control \textbf{U}nit  \\
	$ FVM $ & & \textbf{F}inite \textbf{V}olume \textbf{M}ethod  \\
	$ FEM $ & & \textbf{F}inite \textbf{E}lement \textbf{M}ethod  \\
	$ HTC $ & & \textbf{H}eat \textbf{T}ransfer \textbf{C}oefficient \\
	$ IVC $ & & \textbf{I}nlet \textbf{V}alve \textbf{C}losing \\
	$ PDF $ & & \textbf{P}robability \textbf{D}ensity \textbf{F}unction \\
\end{longtable}

\section{References}
\bibliography{library_app}


\end{document}